\documentclass[journal=iecred,manuscript=article]{achemso}

\usepackage{chemformula} 
\usepackage[T1]{fontenc} 
\usepackage{array}
\usepackage{booktabs} 
\usepackage{listings} 
\usepackage{lmodern} 
\usepackage{mathpazo} 
\usepackage{microtype} 
\usepackage{adjustbox}
\usepackage{xcolor}
\usepackage{float} 
\usepackage{geometry} 
\usepackage{natbib} 
\usepackage{setspace} 
\usepackage{xkeyval} 
\usepackage{caption}

\usepackage{multirow}

\usepackage{graphicx}          
\usepackage{verbatim}
\usepackage{amsmath}
\usepackage{amssymb}                               
\usepackage{blindtext}
\usepackage{empheq}
\usepackage{algorithmic}
\usepackage{booktabs, caption, subcaption, subfloat,algorithm,algorithmic, psfrag}
\usepackage{url,units}
\usepackage{xcolor}
\makeatletter
\def\fase#1{\vcenter{\offinterlineskip\vskip1pt
  \ialign{##\cr
          $\left/\raise.7pt\hbox{$#1$}\right.$\cr
          \noalign{\vskip-.3pt}
          $\mkern1.3mu\hrulefill\mkern4mu$\cr}}}
          \def\fasebig#1#2{\vcenter{\offinterlineskip\vskip1pt
  \ialign{##\cr
          $\left/\raise#2pt\hbox{$#1$}\right.$\cr
          \noalign{\vskip-.3pt}
          $\mkern1.3mu\hrulefill\mkern4mu$\cr}}}
\makeatother

\SectionNumbersOn

\author{Andrea Pozzi}
\affiliation{Dipartimento di Ingegneria Industriale e dell'Informazione, \\ University of Pavia, 27100 Pavia, Italy}
\email{andrea.pozzi03@unipv.it}
\author{Gabriele Ciaramella}
\affiliation{Department of Mathematics and Statistics, University of Konstanz,  Germany}
\author{\\Stefan Volkwein}
\affiliation{Department of Mathematics and Statistics, University of Konstanz,  Germany}
\author{Davide M. Raimondo}
\affiliation{Dipartimento di Ingegneria Industriale e dell'Informazione, \\ University of Pavia, 27100 Pavia, Italy}
\title[An \textsf{achemso} demo]
  {Optimal design of experiments for a lithium-ion cell: parameters identification of a single particle model with electrolyte dynamics}

\abbreviations{IR,NMR,UV}

\begin{document}

\begin{tocentry}
\includegraphics[width=8.47cm, height=4.76cm]{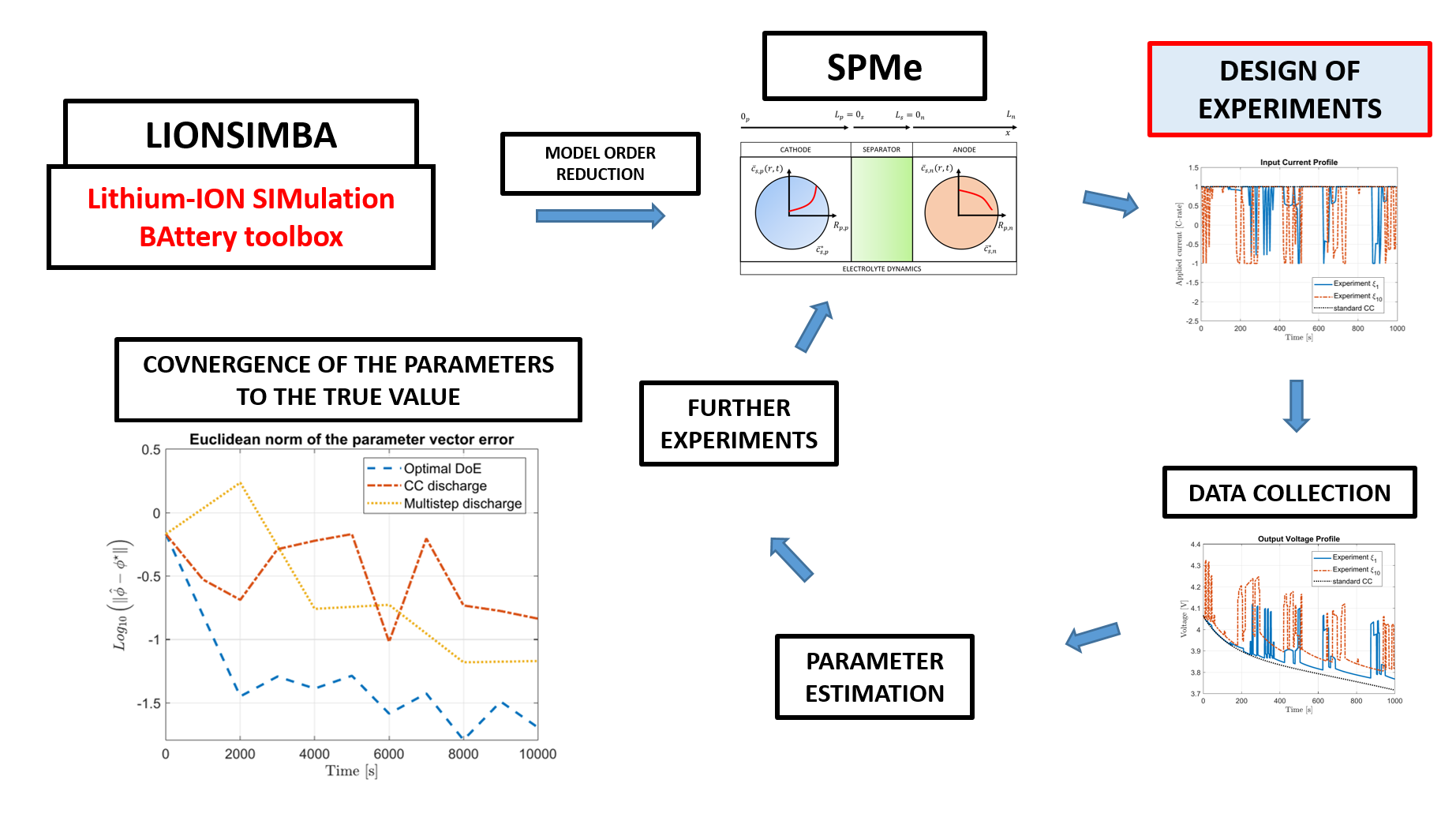} 
\end{tocentry}

\begin{abstract}
Advanced battery management systems rely on mathematical models to guarantee optimal functioning of Lithium-ion batteries. The Pseudo-Two Dimensional (P2D) model  is a very detailed electrochemical  model \textcolor{black}{suitable for  simulations. On the other side,} its complexity prevents its usage in  control and state estimation. \textcolor{black}{Therefore, it is more appropriate the use of simplified electrochemical models such as the Single Particle Model with electrolyte dynamics (SPMe), which exhibits good adherence to real data when suitably calibrated. This work focuses on a Fisher-based optimal experimental design for identifying the SPMe parameters.} The proposed approach relies on a nonlinear optimization  to minimize the  covariance parameters matrix. At first, the parameters are estimated by considering the SPMe as the real plant. Subsequently, a more realistic scenario is considered where the   P2D model  is \textcolor{black}{used to reproduce a real battery behavior.} Results show the effectiveness of the optimal  experimental design when compared to standard  strategies. 
\end{abstract}

\section{Introduction}
Battery Management Systems (BMSs) are necessary in order to provide safe and profitable operations in lithium-ion batteries \cite{chaturvedi2010algorithms}. Advanced BMSs rely on mathematical models whose accuracy is  fundamental   to achieve high performance. Several models have been proposed in literature to describe the behavior of lithium-ion cells. These can be classified in two main categories: Equivalent Circuit Models (ECMs) and Electrochemical Models (EMs). While ECMs are simple and intuitive, EMs, which describe the chemical phenomena occurring inside a cell, can be far  more accurate. The Pseudo Two Dimensional (P2D) model, also known as Doyle-Fuller-Newman (DFN), is a very detailed EM  suitable  for \textcolor{black}{simulating the behavior of Lithium-ion cells} \cite{doyle1993modeling,doyledesign}. \textcolor{black}{Different numerical implementations of the P2D  have been proposed over the years (e.g. DUALFOIL , LIONSIMBA \cite{torchio2016_LIONSIMBA}). The use of simulators to assess the performance of novel strategies is a common procedure in literature. \cite{kovatchev2009silico} However, in order  to obtain  reliable representations of a real process, realistic parameters are required. To this purpose, within the context of Li-ion batteries, Ecker et al.\cite{ecker2015parameterization} performed invasive measurements on a commercial cell. In particular, they obtained a complete parameterization of the DFN model by opening the cell under argon atmosphere. While this approach is suitable for building a realistic simulator, it can not be applied  in the context of battery control since it could  compromise the proper functioning of the cells. On the other side, experiments  performed   under normal operating conditions, based on voltage measurements only, demonstrated that not all the parameters of the P2D model are identifiable.} The structural identifiability and  ill-conditioning sources of the DFN model have been analyzed by many authors (see e.g. Bizeray \cite{bizeray2016state}, where Table 1.1. reports a summary of the literature on the parameter estimation of electrochemical lithium-ion battery models). In Lopez et al. \cite{lópez2016computational} for example, the authors rely on the Fisher Information Matrix \cite{akaike1998information}, which  gives a measure of how much a certain input signal is informative  in terms of parameter sensitivity. The authors show that multiple discharging experiments with several  currents may improve parameters identification, reducing their variances.  However, some parameters of the DFN  still remain  unidentifiable. Similar results have been obtained by Forman et al. \cite{forman2012genetic}, where  the Fisher Information has been used in order to evaluate the accuracy of the DFN parameters estimated using a genetic algorithm.  In Zhang et al. \cite{zhang2014parameter} a sensitivity analysis of up to 30 parameters of the P2D model is conducted and a step-wise experiment  is proposed. In particular, the parameters with similar identifiability conditions are  estimated in the same step, while assuming the others to be known. The proposed approach exploits genetic algorithm for parameters estimation and provides significant results, although it is validated on synthetic data in absence of measurement error. Note that, even in this favorable scenario it was not possible to identify all the parameters.
These results, together with the lack of observability of the P2D model \textcolor{black}{\cite{moura2015estimation}},  motivate the use of simpler models for control purposes.  
In Lu et al. \cite{lu2013review}   an overview of different simple battery models and parameter identification techniques is presented, while in He et al.\cite{he2011evaluation} and Tian et al. \cite{tian2017parameter} the parameter identification process for an ECM is described. In Torchio et al. \cite{torchio2015real},  a step response model, suitable for control purposes, is identified using the least squares  method. The same authors  propose a piece-wise affine approximations  of the P2D model\cite{torchio2017design}. \textcolor{black}{Despite to their simplicity, the use of ECMs and input-output models  does not allow to take into account physico-chemical phenomena occurring inside the battery. For this reason, the Single Particle Model\cite{santhanagopalan2006review} (SPM), a simplified electrochemical model, has been considered by many authors. In particular, the observability of the SPM has been addressed by Di Domenico et al. \cite{di2010lithium}. In this work the authors claim that the Lithium concentration states of the SPM with two electrodes are weakly observable from the differential voltage measurement alone.  On the other side, the parameters identifiability of the SPM has been discussed by e.g.  Bizeray et al. \cite{bizeray2017identifiability}. In this work, the output voltage has been linearized around an equilibrium and the parameters  grouped into hyper-parameters that can be experimentally identified. } A different simplified electrochemical model has been considered in Schmidt et al.\cite{schmidt2010experiment}, where a \textcolor{black}{Fisher Information} approach in combination with a  sensitivity analysis has been used to estimate the identifiability of the parameters. 

\textcolor{black}{In this work we rely on the  SPM with electrolyte dynamics (SPMe), motivated by the fact that reduced electrochemical models preserve a physico-chemical meaning and provide a good fit. In particular, the authors in   Moura et al. \cite{moura2017battery}  demonstrated the suitability of SPMe for control and estimation purposes. }
Nevertheless, model reduction usually comes with a loss of accuracy, which has to be balanced with a  proper parameter identification procedure. 

In order to obtain high parameters accuracy, a sufficiently exciting input is usually  required. The Design of Experiment (DoE) consists in finding the optimal input sequence able to minimize the uncertainty of  model parameters  \cite{pukelsheim1993optimal}. In Liu et al. \cite{liu2016can} the importance of DoE in the context of control of lithium-ion batteries has been underlined. This topic has been discussed  by many authors. In
Mendoza et al. \cite{mendoza2016optimization}  the  thermal cycle of a cell is optimized   in order to estimate the entropy coefficient while reducing experimental time and maximizing the \textcolor{black}{Fisher Information}. This method has been applied by the same authors   in order to  identify the parameters of an ECM coupled with a thermal model\cite{mendoza2017maximizing}, while the authors in Hametner et al.\cite{hametner2013state} propose an optimal design of experiment for a fuzzy model of a Li-ion cell. The DoE has also been  considered by Mathieu et al.\cite{mathieu2017d} in order to optimally calibrate the parameters of a battery ageing model. 
\textcolor{black}{
In Park et al. \cite{park2018optimal} the optimal DoE for the complete parametrization of the  DFN model is addressed. The parameters are firstly divided into two different sets, accordingly to their characteristic, and  an  optimal input profile is iteratively  selected from an input library in order to maximize the corresponding Fisher Information Matrix. The proposed approach is then experimentally validated. Even in this case, some parameters have not been identified with high accuracy and exhibit a wide confidence interval, thus confirming the need of a simpler model for control purposes.}

\textcolor{black}{In this work we propose  an optimal DoE for the parameters identification of the SPMe, for which  the parameters accuracy  remains  fundamental in order to achieve high performance in the context of control and state estimation}.
In the following, we rely on a \textcolor{black}{Fisher Information}-based approach. Considering that the generalized inverse of the  \textcolor{black}{Fisher Information Matrix} provides an approximation of the covariance parameter matrix, we pose a nonlinear optimization problem aiming to minimize the trace of this latter, similarly to what proposed by Korkel et al. \cite{korkel2005numerical}. \textcolor{black}{Differently from  Park et al. \cite{park2018optimal} we fully  design the input for each time instant, without relying on an a priori  library.}
At first, we consider the SPMe as the real plant, with the output affected by zero mean Gaussian noise. 
 In particular, the accuracy of the estimation is evaluated in terms   of mean and  variance of the parameters.
Within this context, the simulations show the effectiveness of the proposed approach when compared to standard constant current profiles. Results show that the proposed approach can provide a   significant improvement in terms of variance reduction and convergence to the true  values of the parameters in minimum  \textcolor{black}{experimental} time.
\textcolor{black}{Subsequently, we consider a more realistic scenario, in which a real battery simulator based on the P2D model is assumed to be the real process, while the SPMe is used as model for control and estimation. Within this context, the SPMe parameters needs to be estimated with high accuracy on data collected from the P2D model. 
Note that the use of a very detailed model (P2D), together with a numerically stable simulation framework (in this paper LIONSIMBA) and a set of realistic parameters, provides important insights on the real battery behavior and allows to estimate whether an approach (e.g. Design of Experiments) will be worth in practice or not (saving a lot of time during the experimental phase).}
The results highlight that the SPMe, simulated using the  parameters from the optimal DoE, presents the best fitting in terms of  prediction of the P2D output voltage.\\
\textcolor{black}{Summarizing, the main contributions of this work are the following.
\begin{itemize}
\item The use of the optimal DoE approach in order to minimize the covariance of the SPMe estimated parameters and improve the model accuracy. 
\item We propose a  sub-optimal approach so to reduce the computational burden of the nonlinear constrained optimization, which may limit the design of long experiments. 
\end{itemize}
}


The paper is organized as follows. In Sec. \ref{sec:p2d_model} the P2D model is described in detail, while the main equations of the SPMe are recalled in Sec. \ref{sec:spm}. In Sec. \ref{sec:method} the optimization method based on the  Fisher information is presented and in Sec. \ref{sec:case_study} the benefits of the proposed method are highlighted in simulation, when the SPMe is affected by Gaussian zero mean noise. In Sec. \ref{sec:p2d_esperiment} the optimal design of experiment is applied considering the P2D model as the real plant. In  Sec. \ref{sec:conclusions}  the obtained results are resumed.

\section{Pseudo two dimension model}\label{sec:p2d_model}
\label{sec:model} 

In this section, the main equations used to describe the electrochemical behavior of a Lithium-ion cell are presented. In particular we rely on the well known isothermal P2D model, which is described by a  set of nonlinear and tightly coupled Partial Differential and Algebraic Equations (PDAEs)\cite{doyledesign}. 

A Lithium-ion cell is composed by a superposition of different layers:  the cathode ($p$), the separator ($s$), and the anode ($n$). The electrodes and the separator are immersed in an electrolytic solution, thus enabling ionic conduction. 
In the following, the index $i \in \{p, s, n\}$ is used to refer to the different  sections of the battery, whose thicknesses
are denoted by $L_i$. 
The diffusion process within the solid phase of each electrode  is described by Fick's law
\begin{align} \label{eq:PDE_p2d}
\frac{\partial c_{s,i}(x,r,t)}{\partial t}=\frac{1}{r^2}\frac{\partial}{\partial r} \left[D_{s,i} r^2 \frac{\partial c_{s,i}}{\partial r}(x,r,t)\right],
\end{align}
where $t \in \mathbb{R}^{+}$ represents the time, $x \in \mathbb{R}$ is the one-dimensional spatial variable along which ions flow, $r\in \mathbb{R^+}$ is the radial direction along which the ions intercalate within the active particles (i.e. the pseudo-second dimension of the model),   $c_{s,i}(x,r,t)$ is the lithium concentration in the solid phase, $D_{s,i}$ is the solid diffusion coefficient.
The electrolyte diffusion of ions is given by the following  equation
\begin{align}
\epsilon_i \frac{\partial}{\partial t}c_{e,i}(x,t) = \frac{\partial}{\partial x} \left[ D_{\textrm{eff,i}} \frac{\partial c_{e,i}(x,t)}{\partial x}\right] + a_i (1-t_+) j_i(x,t),
\end{align}
where $c_{e,i}(x,t)$ is the electrolytic ion concentration and $j_i(x,t)$ is the ionic flux. Moreover,  $t_+$ defines the transference number and $a_i$ is the specific active surface area defined as
\begin{align}
a_i=\frac{3 (1-\epsilon_{f,i}-\epsilon_i)}{R_{p,i}},
\end{align}
where $\epsilon_{f,i}$ is the filler fraction, $\epsilon_i$ is the electrolyte porosity, $R_{p,i}$ is the particle radius, while $D_{\textrm{eff,i}}$ accounts for the effective diffusion coefficients in the electrolyte. In particular, this latte according to Bruggeman's theory is given by
\begin{align}
D_{\textrm{eff,i}}=\epsilon_i^p D_e,
\end{align}
where $D_e$ is the electrolyte diffusion coefficient, assumed to be constant in $c_e$, and $p$ is the Bruggeman coefficient. 
The ionic flux  is modeled by a Butler-Volmer equation 
\begin{align}
j_i(x,t) =& \begin{cases} 2 i_{0,i} \sinh \left[\frac{0.5 F}{\textrm{R} T} \eta_i(x,t) \right],\,\, & i\in \{p,n\}\\
0,\,\, & i\in \{s\}
\end{cases}
\end{align}
with value  zero within the separator domain, where $\eta_i(x,t)$ represents the electrode overpotential, $T$ is the cell temperature, assumed to be constant, $R$ and $F$ are the universal gas constant and the Faraday's constant, respectively. The intercalation exchange current density $i_{0}$ is defined as
\begin{align}
i_{0,i}=& k_{\textrm{i}} \sqrt{c_{e,i}(x,t)(c_{s,i}^{\textrm{max}}-c_{s,i}^*(x,t))c_{s,i}^*(x,t)},
\end{align}
where $c_{s,i}^{\textrm{max}}$ is the maximum allowed concentration in each electrode, $c_{s,i}^{*}(x,t)$ denotes the Li-ions surface concentration, and $k_{\textrm{i}}$ the  kinetic reaction rate. 
The solid phase potential  $\Phi_{s,i}(x,t)$ inside the two electrodes is modeled according to the Ohm's law
\begin{align}
\frac{\partial}{\partial x}\left[\sigma_{\textrm{eff,i}} \frac{\partial}{\partial x} \Phi_{s,i}(x,t) \right] = a_i F j_i(x,t),
\end{align}
where $\sigma_{\textrm{eff,i}}$ is the electrodes effective conductivity
\begin{align}
\sigma_{\textrm{eff,i}}=\epsilon_i^p \sigma_i
\end{align}
with $\sigma_i$  the electrodes  conductivity.
Similarly, Ohm's law is also used for the electrolytic potential $\Phi_{e,i}(x,t)$ 
\begin{align}
a_iFj_i(x,t) +\frac{\partial}{\partial x}\left[\kappa_{\textrm{eff,i}}(c_{e,i}(x,t)) \frac{\partial}{\partial x}\Phi_{e,i}(x,t) \right]= 
\dfrac{\partial}{\partial x}\left[\frac{2 \kappa_{\textrm{eff,i}}(c_{e,i}(x,t)) \textrm{R} T}{F}(1-t_+)\frac{\partial}{\partial x}\ln c_{e,i}(x,t)\right]&.
\end{align}
In particular, $\kappa_{\textrm{eff,i}}(c_{e,i}(x,t))$ is the effective conductivity of the liquid phase
\begin{align}
\kappa_{\textrm{eff,i}}(c_{e,i}(x,t))=\epsilon_i^p \kappa(c_{e,i}(x,t)),
\end{align}
where $\kappa(c_{e,i}(x,t))$ is the electrolyte conductivity  coefficient, assumed to be a polynomial function of the  electrolyte concentration as in Lopez et al.\cite{lópez2016computational}
\begin{align}
\kappa(c_{e,i}(x,t))=&h_1+h_2s+h_3s^2+h_4s^3+h_5s^4,
\end{align}
with $s=10^{-3} c_{e,i}(x,t)$ and real coefficients $h_i,i=1,\cdots, 5$. Finally, the electrode overpotential is defined as follows
\begin{align}
\eta_i(x,t)=\Phi_{s,i}(x,t)-\Phi_{e,i}(x,t)-U_i(\theta_i(x,t)),
\end{align}
where the  open circuit potential of  each electrode $U_i(\theta_i(x,t))$ is given by a nonlinear polynomial function of the surface stoichiometry $\theta_i(x,t)=\frac{c_{s,i}^*(x,t)}{c_i^{max}}$ as in Lopez et al. \cite{lópez2016computational} 
\begin{subequations}
\begin{align}
U_p(\theta_p)=& f_1 +f_2 \tanh{\left(f_3\theta_p +f_4\right)}+\frac{f_5}{(f_6-\theta_p)^{f_7}}+f_5 f_8+f_9e^{f_{10}\theta_p^{f_{11}}}+f_{12}e^{f_{13}(\theta_p+f_{14})}
\\
U_n(\theta_n)=& g_1+g_2e^{g_3\theta_n}+g_4e^{g_5\theta_n}.
\end{align}
\end{subequations}

The cell output voltage is given by
\begin{align}
V_{\textrm{out}}(t) = \Phi_{s,p}(0_p,t) - \Phi_{s,n}(L_n,t).
\end{align}
The constants, the model parameters  and the  coefficients of  $U_p$, $U_n$ and $\kappa_i$ are taken from Lopez et al. \cite{lópez2016computational}, except for  the cathode diffusion coefficient $D_{s,p}$ and the electrolyte diffusion coefficient $D_e$  which are taken from Ecker et al. \cite{ecker2015parameterization}. In particular, the  diffusion coefficients are chosen to better approximate real lithium-ion cell behavior.
For more details about the P2D model and boundary conditions, the reader can refer to Ramadass et al. \cite{Ramadass2004} and Northrop et al.\cite{northrop2011coordinate}. 
In the following, we rely on the numerical implementation of the P2D model provided by the freely available Li-ION SImulation BAttery Toolbox (\textit{LIONSIMBA}\cite{torchio2016_LIONSIMBA}), using the parameters as discussed above.

\section{Single particle model with electrolyte dynamics}\label{sec:spm}
The P2D model is \textcolor{black}{a very detailed model}, but also very complex to be used within the context of battery state estimation and  control. For this reason, in the following model simplifications of the P2D  are   considered. The SPM\cite{santhanagopalan2006review} has been   used  by many authors in the context of battery state estimation  and control  \cite{moura2012pde,tanim2015state,moura2017battery,chaturvedi2010modeling}. The model is obtained from the P2D by approximating the solid phase of each electrode with a single spherical particle. Such particle presents an equivalent area equal  to the one of the  solid phase in the porous electrode. In the SPM basic formulation,  the diffusion of the electrolyte concentration  and the thermal effects are assumed negligible.  The diffusion of the ion concentration is approximated by its average  along the $x$ axis $\overline{c}_{s,i}(t,r_i)$, with the following equation
\begin{align}\label{eq:PDE}
\frac{\partial \overline{c}_{s,i}(r,t)}{\partial t}=\frac{D_{s,i}}{r^2} \frac{\partial}{\partial r} \left( r^2 \frac{\partial \overline{c}_{s,i}(r,t)}{\partial r} \right).
\end{align}
The input current $I_{app}(t)$ enters the model in the Neumann boundary conditions
\begin{align} 
\left. \frac{\partial \overline{c}_{s,i}(r,t)}{\partial r}\right|_{r=0}=0, \hspace{1cm} D_{s,i}\left.\frac{\partial \overline{c}_{s,i}(r,t)}{\partial r}\right|_{r=R_{p,i}}=-j_i,
\end{align}
where $j_p=-\frac{I_{app}(t)}{a_pL_p F A}$ and $j_n=\frac{I_{app}(t)}{a_nL_n F A}$. The initial condition  is given by
\begin{align}
\overline{c}_{s,i}(r,0)=\overline{c}^0_{s,i}(r),
\end{align} 
where $\overline{c}^0_{s,i}(r)$ is the initial concentration profile over the radial axis. 
Due to its simplicity, the SPM results particularly suitable for state estimation and control purposes. However, it shows inaccuracy for current greater than $0.5 I_{1C}$, in particular in case of low electrolyte conductivity \textcolor{black}{as discussed by Moura et al.\cite{moura2017battery}. In that work a single particle model with electrolyte dynamics (SPMe) is proposed} in order to increase voltage prediction accuracy while maintaining the computational effort at a reasonable level. The same electrolyte diffusion equations as in Moura et al. \textcolor{black}{\cite{moura2017battery}} are here adopted 
\begin{subequations}\label{eq:electrolyte_PDE}
\begin{align}
\epsilon_p \frac{\partial  c_{e,p}(x,t)}{\partial t}&= D_{eff,p} \frac{\partial^2 c_{e,p}(x,t)}{\partial x^2}  -\frac{1-t_+}{FAL_p}I_{app}(t)\\
\epsilon_s \frac{\partial  c_{e,s}(x,t)}{\partial t}&= D_{eff,s}  \frac{\partial^2 c_{e,s}(x,t)}{\partial x^2}\\
 \epsilon_n \frac{\partial c_{e,n}(x,t)}{\partial t}&= D_{eff,n}  \frac{\partial^2 c_{e,n}(x,t)}{\partial x^2}+\frac{1-t_+}{FAL_n}I_{app}(t)
\end{align}
\end{subequations}
for $(x,t) \in (0,L) \times (0,T)$. The boundary conditions are given by
\begin{subequations}
\begin{align}
\frac{\partial  c_{e,p}(0_p,t)}{\partial x}&= \frac{\partial  c_{e,n}(L_n,t)}{\partial x}=0\\
 D_{eff,p}  \frac{\partial  c_{e,p}(L_p,t)}{\partial x}&=  D_{eff,s}  \frac{\partial  c_{e,s}(0_s,t)}{\partial x}\\
 D_{eff,s}  \frac{\partial  c_{e,s}(L_s,t)}{\partial x}&=  D_{eff,n}  \frac{\partial  c_{e,n}(0_n,t)}{\partial x}
\end{align}
\end{subequations}
and the initial condition is 
 \begin{align}
c_{e,i}(x,0)=c^0_{e,i},
\end{align} 
where $c^0_{e,i}$ is the initial electrolyte concentration in the  layer $i$, with $i\in \{p,s,n\}$. 
\subsection{Further simplification of the SPMe} \label{sec:SPM_simplified}
\begin{figure}[h]
\begin{center}
\includegraphics[width=0.5\textwidth]{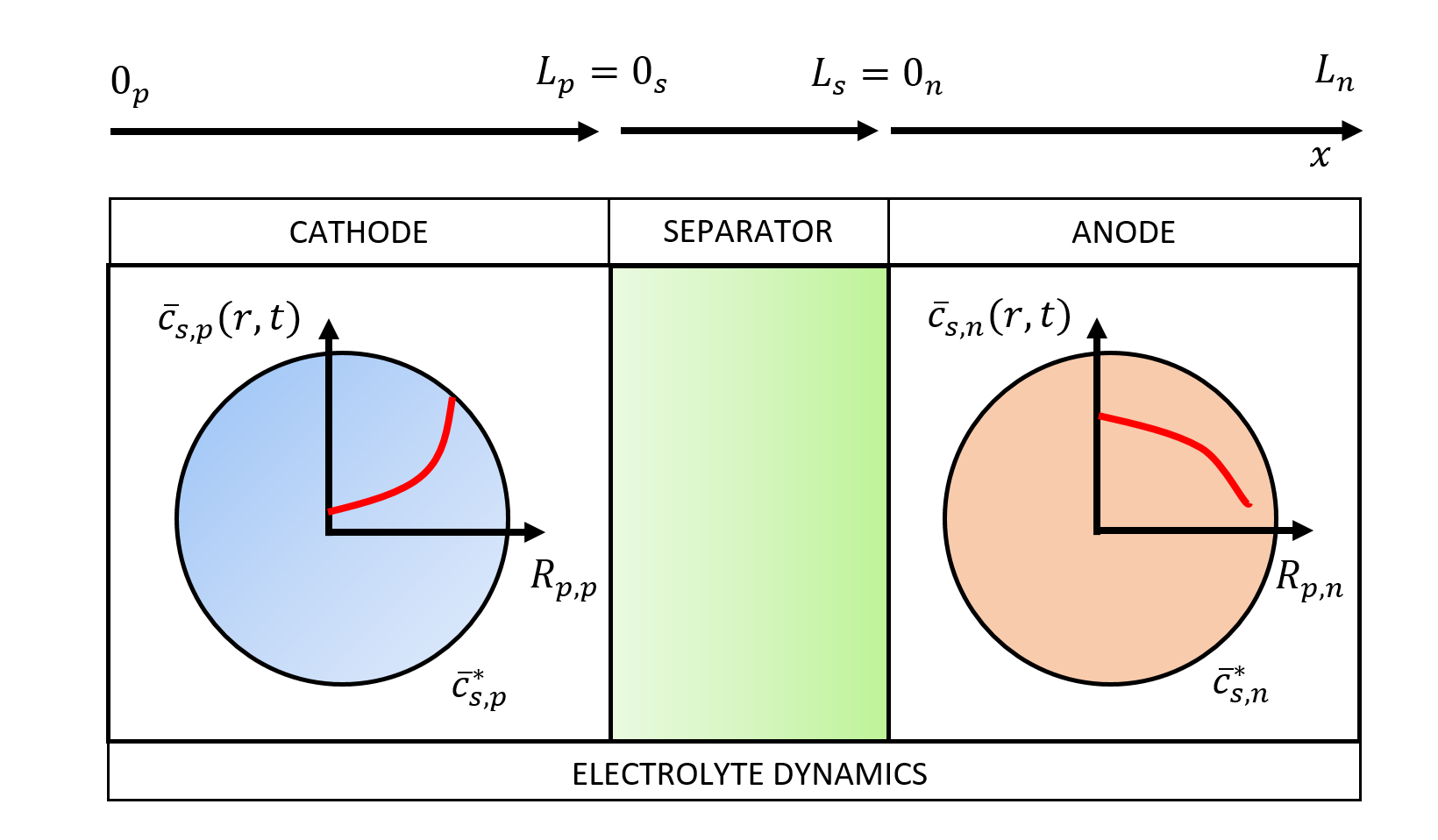}    
\caption{\textcolor{black}{Schematic representation of the SPMe according to Subramanian et al. \textcolor{black}{\cite{subramanian2005efficient}}}}  
\label{fig:SPMe_scheme}                                 
\end{center}                                 
\end{figure}

\begin{figure}[htpb]
\begin{center}
\includegraphics[width=0.5\textwidth]{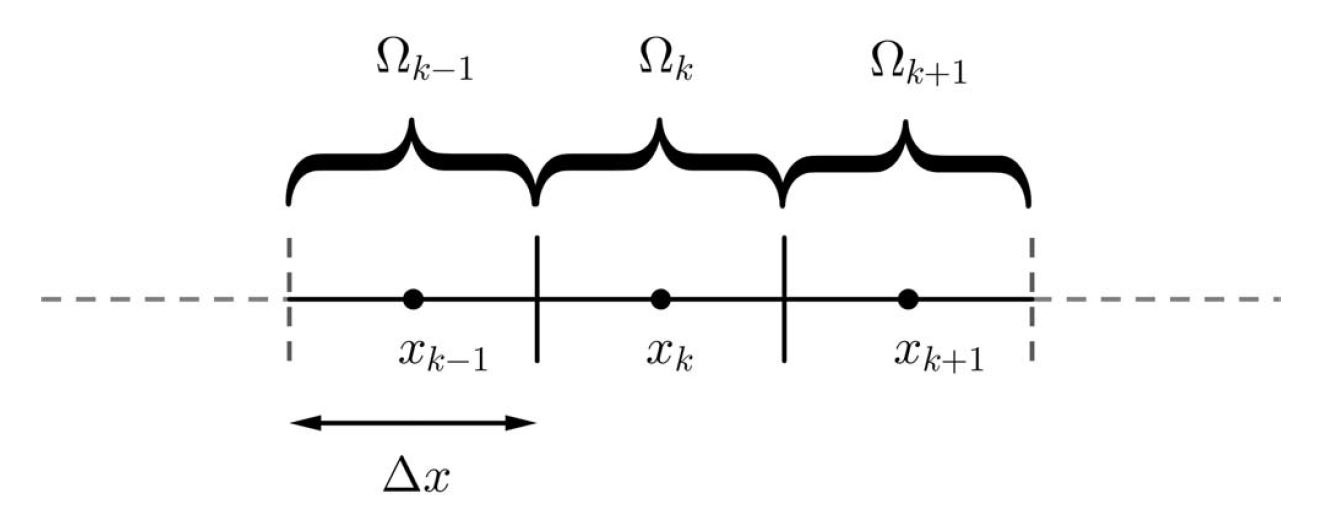}    
\caption{Finite volume discretization.}  
\label{fig:finite_volume}                                 
\end{center}                                 
\end{figure}

In this section,  the partial differential equation  \eqref{eq:PDE} which describes the lithium-ions diffusion in the solid phase is approximated by a set of Ordinary Differential Equations (ODEs). \textcolor{black}{The solid diffusion dynamics can be reduced to a lower order system in several ways. In Bizeray et al. \cite{bizeray2015lithium}  the diffusion dynamic in the solid phase  is spatially discretised using a  Chebyshev orthogonal collocation  method which enables fast and accurate simulations.}
In Subramanian et al. \cite{subramanian2005efficient}  the lithium concentration profile in the particles is approximated as a polynomial function (see Figure \ref{fig:SPMe_scheme}) 
\begin{align}
{c}_{s,i}(r,t)=a(t)+b(t)\frac{r^2}{R^2_{p,i}}+d(t)\frac{r^4}{R^4_{p,i}},
\end{align} 
where the coefficients $a(t)$, $b(t)$, $d(t)$ are expressed in terms of  volume-averaged lithium concentration $\overline{c}^{avg}_{s,i}(t)$,  volume-averaged concentration flux $q^{avg}_i(t)$ and surface concentration $\overline{c}^{*}_{s,i}(t)$. In this work, the model is simplified according to Subramanian et al. \textcolor{black}{\cite{subramanian2005efficient}} and   Fick's law is reduced into a set of $N^{ode}=4$ ODEs 
\begin{subequations}\label{eq:diff_states}
\begin{align} 
\dot{\overline{c}}^{avg}_{s,p}(t)&= \frac{3}{R_{p,p}FAL_pa_p}I_{app}(t)\\
\dot{\overline{c}}^{avg}_{s,n}(t)&= -\frac{3}{R_{p,n}FAL_na_n}I_{app}(t)\\
\dot{q}^{avg}_p(t)&=-30 \frac{D_{s,p}}{R^2_{p,p}}q^{avg}_p(t)+\frac{45}{2R^2_{p,p}}\frac{1}{FAL_pa_p}I_{app}(t)\\ 
\dot{q}^{avg}_n(t)&=-30 \frac{D_{s,n}}{R^2_{p,n}}q^{avg}_n(t)-\frac{45}{2R^2_{p,n}}\frac{1}{FAL_na_n}I_{app}(t).
\end{align}
\end{subequations}
The positive and negative surface concentrations are then given by
\begin{subequations}
\begin{align}
\overline{c}^*_{s,p}(t)&=\overline{c}^{avg}_{s,p}(t)+\frac{8R_{p,p}}{35}q^{avg}_p(t)+\frac{R_{p,p}}{35D_{s,p}}\frac{1}{FAL_pa_p}I_{app}(t)\\
\overline{c}^*_{s,n}(t)&=\overline{c}^{avg}_{s,n}(t)+\frac{8R_{p,n}}{35}q^{avg}_n(t) -\frac{R_{p,n}}{35D_{s,n}}\frac{1}{FAL_na_n}I_{app}(t).
\end{align}
\end{subequations}
The output voltage  is a nonlinear function of  the states and the input
\begin{align}
V(t)&=U_p(\overline{\theta}_p(t))-U_n(\overline{\theta}_n(t)) + \Delta \Phi_e(t)+ \overline{\eta}_p(t)-\overline{\eta}_n(t),
\end{align}
where  $\overline{\theta}_p(t)$ and $\overline{\theta}_n(t)$ are the positive and negative  average surface stoichiometries, respectively given by
\begin{align}
\overline{\theta}_p(t)=\frac{\overline{c}^{*}_{s,p}(t)}{c_p^{max}}, \hspace{1cm}
\overline{\theta}_n(t)=\frac{\overline{c}^{*}_{s,n}(t)}{c_n^{max}}.
\end{align}
$\Delta \Phi_e(t)$ is the average electrolyte potential drop, which is approximated as
\begin{align}
\Delta \Phi_e(t)=&-\frac{I_{app}}{2A}\left(\frac{L_p}{\overline{\kappa}_{eff,p}}+\frac{2L_s}{\overline{\kappa}_{eff,s}}+\frac{L_n}{\overline{\kappa}_{eff,n}}\right)+\frac{2RT}{F}(1-t_+)\ln{\frac{c_e(0_p,t)}{c_e(0_n,t)}},
\end{align}
where $\overline{\kappa}_{eff,i}$ is the average effective electrolyte conductivity  
\begin{align}\label{eq:electrolyte_conductivity}
\overline{\kappa}_{eff,i}=\epsilon_i^p \kappa(\overline{c}_{e,i}),
\end{align}
with  $\overline{c}_{e,i}= \frac{1}{L_i}\int_{L_i} c_{e,i}(x,t)\,dx$  the average electrolyte concentration in layer $i$, with $i \in \{p,s,n\}$. Note that, differently from Moura et al.\cite{moura2017battery}, in which the electrolyte conductivity is assumed uniform in the electrolyte concentration ($\kappa(c_e)\simeq \overline{\kappa}$), in this work we adopt a more realistic approximation. \textcolor{black}{In particular, the effective electrolyte conductivity $\overline{\kappa}_{eff,i}$ is evaluated according to \eqref{eq:electrolyte_conductivity}  using for each section the different average value of the electrolyte concentration  $\overline{c}_{e,i}$. }
Finally, the positive and negative average overpotentials are given by 
\begin{subequations}
\begin{align}
\overline{\eta}_p(t)&=\beta \sinh^{-1} \left(  \frac{-I_{app}(t)}{2AFL_pa_p\overline{i}_{0,p}(t)} \right)\\
\overline{\eta}_n(t)&=\beta \sinh^{-1} \left(  \frac{I_{app}(t)}{2AFL_na_n\overline{i}_{0,n}(t)} \right),
\end{align}
\end{subequations}
where $\beta=\frac{2RT}{F}$ and 
\begin{align}
\overline{i}_{0,i}(t)=k_i\sqrt{\overline{c}_{e,i}(t) \, \overline{c}^{*}_{s,i}(t) \left(c^{max}_i-\overline{c}^{*}_{s,i}(t)\right)}.
\end{align}
The obtained model consists of four ODEs and three Partial Differential Equations (PDEs). The PDEs are discretized using the finite-volume method described in Torchio et al. \cite{torchio2016_LIONSIMBA}, where the spatial domain is divided into $3 \cdot N^{el}$ non-overlapping volumes with centered nodes, as shown in Figure \ref{fig:finite_volume} and $\overline{c}^k_{e,i}(t)$ is the volume average electrolyte concentration over the $k$-th volume.

\section{Optimal design of experiment and parameter estimation method} \label{sec:method}
The value of the physical parameters of a lithium-ion cell varies significantly from cell to cell. This happens even when cells of the same chemistry and the same type are considered. In the context of state estimation and control, the model accuracy is fundamental  in order to achieve high performance. For this reason, rather than using the mean values  of the parameters distribution as in  Ecker et al. \cite{ecker2015parameterization}, it is much better to resort to a parameters identification procedure.  
However, the use of standard \textcolor{black}{identification} current profiles \textcolor{black}{(e.g. constant current, multistep discharging current)} may not be sufficiently informative for this purpose. This has been shown in the context of lithium-ion batteries in, e.g. Lopez et al. \cite{lópez2016computational}, where the authors investigate  the structural identifiability of  the model parameter vector $\phi \in \mathbb{R}^{N_\phi}$
\begin{align}
\phi=\left[p\,\,\,t_+\,\,\,D_e\,\,\,D_{s,p}\,\,\,D_{s,n}\,\,\,k_p\,\,\,k_n \right]^T.
\end{align}
\textcolor{black}{when constant discharging currents are applied. 
The results underline that only a subset of the previous parameter vector is identifiable if the estimation process is based on voltage measurements only. In particular, the cathodic diffusion constant $D_{s,p}$ and the reaction rate constants ($k_p$, $k_n$) remain unidentifiable after multiple experiments.
In this work we focus on the design of experiments to be conducted on a Li-ion cell in order to estimate the same parameter vector $\phi$  as in Lopez et al. \textcolor{black}{\cite{lópez2016computational}}  The objective is to find an optimal input current profile aiming to maximize the accuracy of the vector $\phi$.  We consider $\phi^\star$ as the true value of the parameter vector and $\phi_0$ as the initial guess of the parameter estimation procedure (see Table \ref{tab:parameters_estimated}). Note that, $\phi_0$ has been set equal to the most probable  value of the parameters.}

\begin{table} [htpb] 
\begin{center}
{\renewcommand\arraystretch{1.3} 
\begin{adjustbox}{width={\textwidth},totalheight={\textheight},keepaspectratio}
\begin{tabular}{*5c} 
\toprule
Parameter  								& Unit 							& 	Description 							&	Initial guess $\phi_0$ 	& True value $\phi^\star$ 	\\
\midrule
$p$ 									& -								&	Bruggeman coefficient 					& 	\textcolor{black}{1.6613} 					& $1.5$   					\\
$t_+$ 									& -								&	Cationic transference number 			&	\textcolor{black}{0.4975 }					& $0.363$ 				 	\\
$D_e$ \cite{ecker2015parameterization}									& $\text{m}^2 \text{s}^{-1}$ 	& Diffusion coefficient in the electrolyte 	& \textcolor{black}{$1.3376\cdot 10^{-10}$} 	& $2.44 \cdot 10^{-10}$     	\\
$D_{s,p}$\cite{ecker2015parameterization} 						& $\text{m}^2 \text{s}^{-1}$	& Diffusion coefficient in the cathode 		& \textcolor{black}{$7.98\cdot 10^{-11}$}  	& $7.5\cdot 10^{-11}$    	\\
$D_{s,n}$ 								& $\text{m}^2 \text{s}^{-1}$	& Diffusion coefficient in the anode 		& \textcolor{black}{$1.17 \cdot 10^{-13}$} 	& $10^{-13}$   \\
$k_p$ 									& $ \text{m}^{2.5}\text{mol}^{-0.5} \text{s}^{-1}$ & Kinetic reaction rate constant in the cathode& \textcolor{black}{$1.8266\cdot 10^{-11}$} 	& $2\cdot 10^{-11}$ \\
$k_n$ 									& $ \text{m}^{2.5}\text{mol}^{-0.5} \text{s}^{-1}$ & Kinetic reaction rate constant in the anode & \textcolor{black}{$1.4769\cdot 10^{-11}$} & $2\cdot 10^{-11}$  \\
\bottomrule
\end{tabular}
\end{adjustbox}}
\end{center}
\caption{Initial guess and true value of the parameters to be estimated.}\label{tab:parameters_estimated}
\end{table}

\textcolor{black}{The set of ODEs and  nonlinear output equation describing the SPMe can be expressed  as}
\begin{subequations}\label{eq:dyn_model}
\begin{align} 
\dot{x}(t)&=f(x(t),u(t),\phi) \label{eq:dyn_model1}\\
y(t)&=g(x(t),u(t),\phi) \label{eq:dyn_model2}\\
x(t_0)&=x_0,
\end{align}
\end{subequations}
where the model input $u(t) \in \mathbb{R}$ is the applied current $I_{app}(t)$, the model output  $y(t) \in \mathbb{R}$ is the voltage $V(t)$  and the differential state vector $x(t) \in \mathbb{R}^{m}$, with $m=N^{ode}+3\cdot N^{el}$, is   given by
\begin{align}
x= [& \overline{c}^{avg}_{s,p} \,\,\, \overline{c}^{avg}_{s,n}\,\,\, q^{avg}_p \,\,\, q^{avg}_n\,\,\, x_{\overline{c}_e}]^T,
\end{align}
where 
\begin{align}
x_{\overline{c}_e}=\left[\overline{c}_{e,p}^1\,\,\, ...\,\,\,\overline{c}_{e,p}^{N^{el}}\,\,\, \overline{c}_{e,s}^1\,\,\, ...\,\,\,\overline{c}_{e,s}^{N^{el}}\,\,\,\overline{c}_{e,n}^1\,\,\, ...\,\,\,\overline{c}_{e,n}^{N^{el}}\right]^T.
\end{align}
Moreover, $\phi \in \mathbb{R}^{N_{\phi}}$ is the set of parameters we aim to estimate, $t_0 \in \mathbb{R}^+$ is the initial time and  $x_0 \in \mathbb{R}^{m}$  is the initial state vector.
In the following, we assume that both inputs and outputs are applied/measured every $t_s$ seconds. Let $\mathbf{u}^{\xi} \in \mathbb{R}^{N}$ denote the input sequence for a given experiment $\xi$ over a time interval  $[t^\xi_i,t^\xi_f]$
\begin{align}
\mathbf{u}^{\xi}=\left[u(t^\xi_i), u(t^\xi_i+t_s) ..., u(t^\xi_f)\right],
\end{align}
where the number of control inputs and acquired outputs is $N=\frac{t^\xi_f-t^\xi_i}{t_s}$. Note that the sequence $\mathbf{u}^{\xi}$ corresponds to a piecewise constant input.
Let $\mathbf{y}^{\xi}(\phi) \in \mathbb{R}^{N}$ be the  output  sequence corresponding to the application of the input sequence $\mathbf{u}^{\xi}$ for the given experiment $\xi$, that is
\begin{align}
\mathbf{y}^{\xi}=\left[y(t^\xi_i), y(t^\xi_i+t_s) ..., y(t^\xi_f)\right].
\end{align}
In the following, the output measurements are assumed to be affected by zero mean Gaussian noise, uncorrelated over  time. In particular, for the true value of the parameter vector  $\phi^*$, one has
\begin{align}
\overline{y}(t)&=g(x(t),u(t),\phi^\star) +v(t),
\end{align}
where $v(t)\sim \mathcal{N} (0,\sigma_y^2)$. The vector of the noisy observed data $\mathbf{\overline{y}}^{\xi} \in \mathbb{R}^N$  is defined as follows
\begin{align}
\mathbf{\overline{y}}^{\xi}=\left[\overline{y}(t^\xi_i), ..., \overline{y}(t^\xi_f)\right].
\end{align}
It holds that $\mathbf{\overline{y}}^{\xi}  \sim \mathcal{N} (\mathbf{y}^{\xi}(\phi^\star),C_y)$, where $C_y \in \mathbb{R}^{N \times N}$ is the  measurement covariance matrix. In particular,  $C_y$ is a diagonal matrix with entries given by the measurement error variances $\sigma_y^2$, i.e. 
\begin{align}
C_y= \sigma_y^2 \mathbb{I}_{N},
\end{align}
where $I_N$ is the identity matrix of order $N$. 

\subsection{Structural identifiability, sensitivity matrix and ill-conditioning analysis}\label{sec:ill_conditioning_sources}
The analysis of ill-conditioning sources  in the parameter sensitivity matrix allows to assess the structural identifiability  of  model parameters \cite{bellman1970structural,cobelli1980parameter}. The sensitivity matrix  for a given experiment $\xi$ is given by the Jacobian matrix of the output vector $\mathbf{y}^\xi(\phi)$ with respect to the parameters we aim to estimate, i.e.
\begin{align}\label{eq:sensitivity}
S^\xi(\phi)=\nabla_{\phi} \,\mathbf{y}^\xi(\phi),
\end{align}
with $S^\xi(\phi) \in \mathbb{R}^{N \times N_{\phi}}$. Note that, the sensitivity matrix is not related to measurement data and measurement noise, but depends only on the model equations, on the input vector (i.e. the experiment) and on the parameter vector $\phi$ in which it is evaluated.  For this reason, it is possible  to compute the sensitivity matrix for a particular  experiment $\xi$   directly in simulation. 
 In the following,  the Jacobian matrix of the output vector with respect to the parameters, i.e. the sensitivity matrix, is numerically computed using  the finite difference method. This provides a useful approximation for the columns of the sensitivity matrix, given  as follows
\begin{align}
S^\xi_j=\frac{\mathbf{y}^\xi(\{\phi^1,...\phi^i+h,...\phi^{N_{\phi}}\})-\mathbf{y}^\xi(\phi)}{h}, \hspace{1cm} j=1,2,\, ...,\,N_{\phi},
\end{align}
where $h$ is suitably chosen (in the following  $h=0.001$).
This choice is motivated by the fact that \textcolor{black}{the evaluation of the analytical expression of $S^\xi(\phi)$} becomes prohibitive in case of large dimensions of the output sequence, that is required especially for long horizons problems.  

The finite difference approach in the context of sensitivity analysis is also suggested in the guideline of \textit{CasADi}\cite{andersson2012}, the toolbox we will adopt for optimization purposes (see Sec. \ref{sec:optimization}). 

The ill-conditioning of $S^\xi(\phi)$ implies that one or more parameters are unidentifiable for a given experiment $\xi$. This can be assessed by analizing the singular values $\zeta_1\,...\,\zeta_{N_\phi}$ of the sensitivity matrix\cite{belsley2005regression,barz2015nonlinear,lópez2016computational}, which are computed with  the singular value decomposition  method \cite{golub1970singular}.
In particular, we consider the  condition number 
\begin{align}
\kappa(S^\xi(\phi))=\frac{\zeta^{max}}{\zeta^{min}}
\end{align}
and  the collinearity index 
\begin{align}
\gamma(S^\xi(\phi))=\frac{1}{\zeta^{min}}.
\end{align}
where $\zeta^{max}$ and $\zeta^{min}$ are the maximum and the minimum singular  value of the sensitivity matrix $S^\xi(\phi)$.
High values of the  condition number and of the collinearity index indicate that the  sensitivity matrix is ill-conditioned.


\subsection{Fisher information matrix  and covariance of the parameters}
Once the experiment $\xi$ is performed and the output measurement sequence $\mathbf{\overline{y}}^{\xi}$ is collected, the parameters are estimated by solving the  following maximum likelihood  problem 
\begin{align} \label{eq:est}
\hat{\phi}&= \text{arg} \min_{\phi \in [\phi_{m},\, \phi_{M}] }\left[ (\mathbf{\overline{y}}^{\xi}- \mathbf{y}^\xi(\phi))^T (\mathbf{\overline{y}}^{\xi}- \mathbf{y}^\xi(\phi)) \right],
\end{align}
where $\phi_m$ and $\phi_M$ are suitable lower and upper bound that restrict the feasible region of the optimization problem to a  set that is physically meaningful.
Due to the fact that  $\mathbf{\overline{y}}^{\xi}$ is a random variable, one has that also the estimated parameter vector $\hat{\phi}$ is  a random variable, i.e. $\hat{\phi} \sim \mathcal{N} (\phi^\star,C^\xi_\phi)$. Let  $C^\xi_\phi \in \mathbb{R}^{N_\phi \times N_\phi}$ denotes  the  covariance matrix of the parameters, that depends on the experiment $\xi$. The  design of experiment proposed in the following aims to reduce such covariance. In  particular, we rely on the Fisher Information Matrix \cite{akaike1998information}, denoted by $F^{\xi}(\phi) \in \mathbb{R}^{N_\phi \times N_\phi}$, that is a  symmetric matrix giving a measure of how much an experiment is informative and it is  given by
\begin{align}
F^{\xi}(\phi)=S^{\xi}(\phi)^TC_y^{-1}S^{\xi}(\phi),
\end{align}
\textcolor{black}{The generalized inverse of the Fisher Information Matrix gives a lower bound for  the parameter covariance matrix according to the Cramer-Rao bound} \cite{cover2012elements}
\begin{align}
 F^{\xi}(\phi)^{-1} \leq C^{\xi}_\phi.
\end{align}
The \textcolor{black}{Fisher Information Matrix} for a nonlinear system is strictly dependent on the value of the parameter vector $\phi$ in which it is evaluated, as for the sensitivity matrix.
Note that, by applying in simulation  a standard Constant Current (CC) discharging protocol to the SPMe, the resulting sensitivity matrix $S^\xi(\phi)$ is full rank. This result is promising because implies that there exists an input sequence for which the Fisher matrix is positive definite and therefore its inverse matrix is well defined. 

\subsection{Optimization method} \label{sec:optimization}
\begin{figure}[htpb]
\begin{center}
\includegraphics[width=0.35\textwidth]{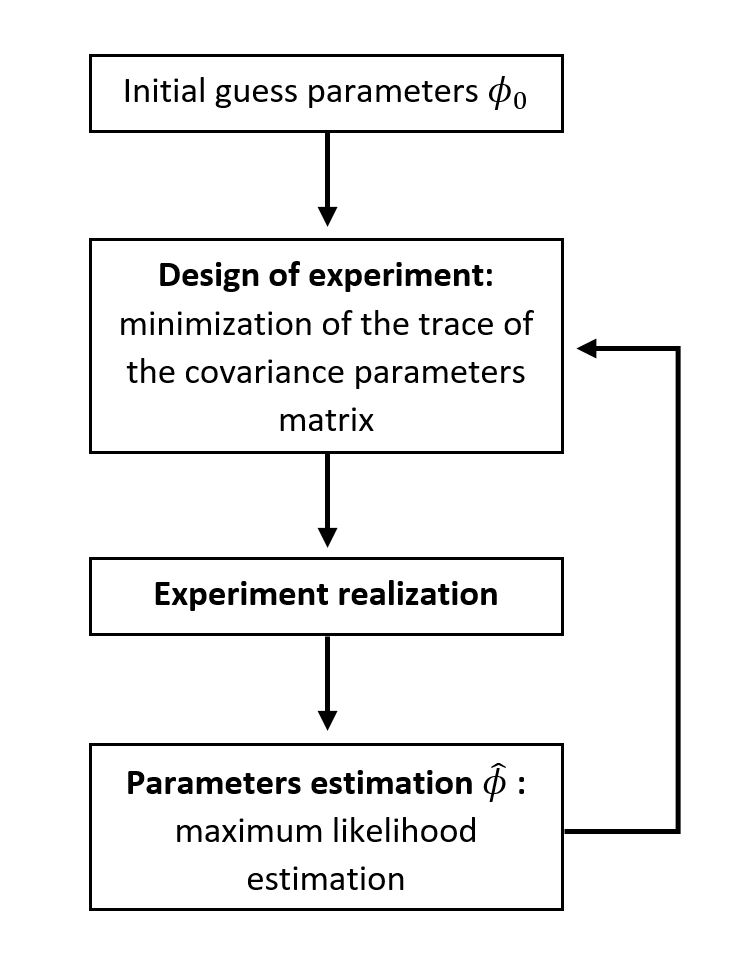}    
\caption{Schematic representation of the optimal experimental design and parameters estimation process.}  
\label{fig:doe_scheme}                                 
\end{center}                                 
\end{figure}

In this section, an iterative optimal design of experiment method is described.
In particular, a sequence of  experiments $\xi_i$, $i=1,\,...\,,n$ is designed in order to  estimate the parameter vector $\hat{\phi}$ with high accuracy and  maximal statistical reliability \cite{korkel2005numerical}.  
\textcolor{black}{At each iteration,  the experiment $\xi_i$  is obtained by minimizing a functional of the parameter covariance matrix $J\left(C^{\xi_i}_{\hat{\phi}_{i-1}}\right)$, i.e. by solving the following constrained optimization problem 
\begin{align} \label{eq:doe_problem}
\min_{\mathbf{u}^{\xi_i}} \begin{matrix} J\left(C^{\xi_i}_{\hat{\phi}_{i-1}}\right) \end{matrix},
\end{align} 
subject to
\begin{subequations}
\begin{align}
\mbox{model dynamics}\\
x(t^{\xi_i}_i)=x^{\xi_i}_0\\
-I_{\text{max}}\leq u(t) \leq I_{\text{max}} \\
h(x(t),u(t),\hat{\phi}_{i-1})\leq 0, \label{eq:safety_constraint}
\end{align}
\end{subequations}
where $I_{\text{max}}$ is a suitable bound for the input and  $x^{\xi_i}_0$ is the initial state for the experiment $\xi_i$.
Since it is not possible to minimize directly the variance of the parameters, several optimization criteria can be used. \textcolor{black}{In the formulation of problem \eqref{eq:doe_problem} we rely on the A-criterion \cite{korkel2004numerical}, i.e. on the minimization of the trace of the approximated covariance  matrix  
\begin{align}
J\left(C^{\xi_i}_{\hat{\phi}_{i-1}}\right) = \text{Tr}\left(C^{\xi_i}_{\hat{\phi}_{i-1}}\right)
\end{align}
}
The resulting input $\mathbf{u}^{\xi_i}$, obtained by solving the above optimization problem, is applied to the model. Subsequently,  the estimated parameter vector is  updated with $\hat{\phi}_i$, obtained  as the solution of the maximum likelihood estimation process \eqref{eq:est} using the output measurements   collected during all the experiments $\xi_1,...,\xi_i$ performed so far. A schematic representation of the method is resumed in Figure  \ref{fig:doe_scheme}. }
Voltage and state of charge constraints are taken into account using the  nonlinear constraint in \eqref{eq:safety_constraint}, that allows to explicitly  consider safety  during the experiments realization. 
In order to guarantee that each experiment has the same initial state, i.e. $x^{\xi_i}_0=x_0$,  a charging procedure, followed by a resting period which brings the system to a steady state, is applied to the cell before each new experiment. 
\textcolor{black}{The maximum likelihood problem in \eqref{eq:est}} as well as the nonlinear optimization problem in \eqref{eq:doe_problem}   have been  solved using \textcolor{black}{the interior point method\cite{wright1999numerical, potra2000interior,wachter2006implementation}} and  \textit{CasADi} \cite{andersson2012}, an open source tool which provides a symbolic framework for   nonlinear numerical optimization algorithms. 
Furthermore, the solver efficiency has been improved by performing a scaling procedure  to the parameters in Table \ref{tab:parameters_estimated} over their nominal values ($\phi^*$), which may vary among several orders of magnitude\cite{korkel2005numerical}. \textcolor{black}{In real battery experiments, the parameters should be scaled considering their likely values, which are indicated on the cell data-sheet.}

\subsection{Computational cost reduction by sub-optimal approach}\label{sec:suboptimal_method}
\begin{figure}[htpb]
\begin{center}
\includegraphics[width=0.35\textwidth]{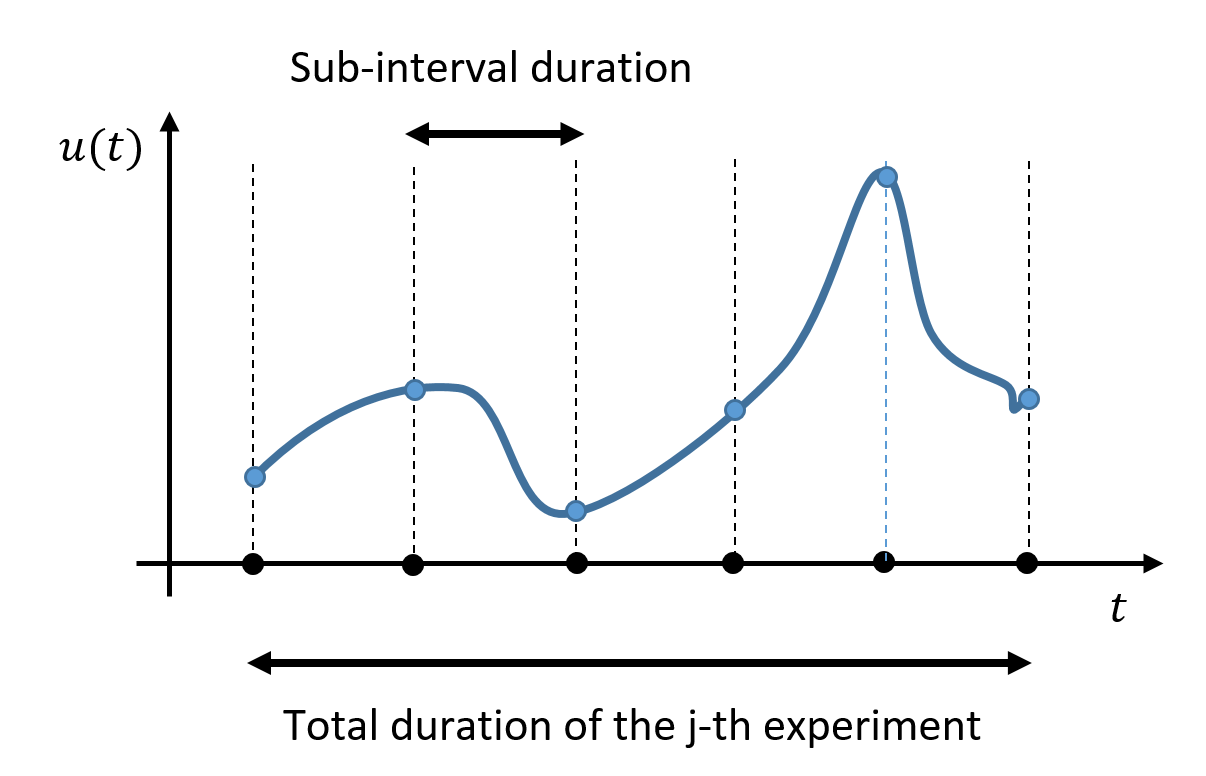}    
\caption{Schematic representation of the sub-optimal experimental design.}  
\label{fig:suboptimal}                                 
\end{center}                                 
\end{figure}

Since the objective of each experiment is to lead to a covariance parameter matrix low enough to guarantee a meaningful parameter estimation, the  input may  need to be optimized over a long experimental time. This, together with the nonlinearities of the model, could increase dramatically the computational burden  of problem \eqref{eq:doe_problem}. For this reason, we divide the design of the experiment $\xi_i$ into $M$ sub-problems,\textcolor{black}{each involving only a fraction of the overall set of optimization variables,} as shown in Fig. \ref{fig:suboptimal}. 
Each step aims to design the input vector $\mathbf{u}^{\xi_{i,j}}\in \mathcal{R}^{\tilde{N}}$, $j=1,\,...\,M$, over the time interval $[t_i^{\xi_{i,j}},t_f^{\xi_{i,j}}]$, where $\tilde{N}=\frac{t^{\xi_{i,j}}_f-t_i^{\xi_{i,j}}}{t_s}$, i.e. $\tilde{N}=\frac{N}{M}$.
\textcolor{black}{At the $j$-th step, the input $\mathbf{u}^{\xi_{i,j}}$ is designed by minimizing the trace of  $C^{\left[ \xi_{i,1}\,...\,\xi_{i,j}\right]}_\phi$, that is the covariance matrix over the $1,\cdots,j$ partial experiments performed so  far in simulation. In particular, at each step the following constrained optimization problem is solved
\begin{align} \label{eq:suboptimal_proble}
\min_{\mathbf{u}^{\xi_{i,j}}} \begin{matrix} \text{Tr}\left(C^{\left[ \xi_{i,1}\,\cdots\,\xi_{i,j}\right]}_{\phi_{i-1}}\right) \end{matrix}
\end{align} 
subject to
\begin{subequations}
\begin{align}
\mbox{model dynamics}\\
x(t^{\xi_{i,j}}_i)=x^{\xi_{i,j}}_0\\
-I_{\text{max}}\leq u(t) \leq I_{\text{max}} \\
h(x(t),u(t),\hat{\phi}_{i-1})\leq 0
\end{align}
\end{subequations}}
The different subproblems are solved in sequence, $j=1,\,...\,M$ and for each $j=2,\,...\,M$, the initial condition is obtained by simulating the effect of the input designed so far on the SPMe starting  with initial condition $x_0$.  Once all the $M$ steps have been performed,   one obtains a sub-optimal input sequence, given by
\begin{align}
\mathbf{\tilde{u}}^{\xi_i}=\left[\mathbf{u}^{\xi_{i,1}},\,\mathbf{u}^{\xi_{i,2}},\,...\, \mathbf{u}^{\xi_{i,M}}\right].
\end{align} 
Only at this point the sequence $\mathbf{\tilde{u}}^{\xi_i}$ is applied to the system and the result of the experiment $\xi_i$ is then obtained. Finally,  $\hat{\phi}_i$ is estimated solving \eqref{eq:est}. The sub-optimality comes from the division of the DoE into $M$ sub-problems, which was required for reducing the complexity of the optimization problem. \textcolor{black}{Thanks to this scheme,   it is possible to consider  long experiments, useful for achieving low  covariance of the parameters,  without dramatically increasing the computational burden.} \textcolor{black}{The optimal input design process ends when one of the following conditions is satisfied:
\begin{itemize}
\item a variance threshold is achieved;
\item a maximum numbers of experiments has been executed;
\item no more significant covariance decrease is obtained.
\end{itemize}}
This allows us  to estimate the parameter vector $\hat{\phi}_i$ with  high accuracy, while maintaining the computational burden at a reasonable level.  In  Algorithm \ref{alg:suboptimal} the main  features of the sub-optimal approach are resumed. 

\begin{algorithm} 
    \caption{Sub-optimal design of experiment $\xi_i$.}
    \label{alg:suboptimal}
\begin{algorithmic}[1]
\STATE Initialize the time: $t_i^{\xi_{i,1}}=t_i^{\xi_{i}}$
\STATE Initialize the state: $x_0^{\xi_{i,1}}=x_0^{\xi_{i}}$
 \FOR{$j=1$ \TO $M$}  
 	\STATE {Compute the input by solving the optimization problem  \eqref{eq:suboptimal_proble}
} 
\STATE {Update the time: $t_i^{\xi_{i,j+1}}=t_i^{\xi_{i,j}}+ \frac{t_f^{\xi_{i}}-t_i^{\xi_{i}}}{M}$}
\STATE {Update the  state applying the input sequence computed so far to  the SPMe in simulation: \\$x_0^{\xi_{i,j+1}}=x\left(t_i^{\xi_{i,j}}+ \frac{(t_f^{\xi_{i}}-t_i^{\xi_{i}})}{M}\right)$}
 \ENDFOR    
 \STATE Concatenate  the sub-problems solutions in order to  obtain the input sequence of the experiment $\xi_i$: \\$\mathbf{\tilde{u}}^{\xi_i}=\left[\mathbf{u}^{\xi_{i,1}}\,...\, \mathbf{u}^{\xi_{i,M}}\right]$
  \STATE Update the estimated parameter vector: \\$\hat{\phi}_i= \text{arg} \min \left[ (\mathbf{\overline{y}}^{\xi_i}- \mathbf{y}^{\xi_i}(\phi_i))^T (\mathbf{\overline{y}}^{\xi_i}- \mathbf{y}^{\xi_i}(\phi_i)) \right]$,\\ where $\phi_i \in [\phi_{m},\, \phi_{M}] $
\end{algorithmic}
\end{algorithm}

In Table \ref{tab:suboptimal_comparison} the sub-optimal approach  is compared with  the optimal one which relies on the numerical solution of  the optimization problem in \eqref{eq:doe_problem} over the whole time horizon $t_f^{\xi_{i}}-t_i^{\xi_{i}}$. The simulation is performed for     $M=4$, $t_f^{\xi_{i}}-t_i^{\xi_{i}}=200\,s$, \textcolor{black}{$t_s=5s$} and $\hat{\phi}_{i-1}=\phi_0$. In particular, the two approaches are compared in terms of computational time and trace of the  covariance matrix. As it can be noticed, the sub-optimal approach presents a significant burden reduction while achieving  high parameters accuracy. 
\begin{table} [!htpb] 
\begin{center}
{\renewcommand\arraystretch{1.4} 
\begin{tabular}{*3c} 
\toprule
 Optimization method & Computational time $[s]$  & Scaled $  \text{Tr}\left(C^{\xi}_\phi\right)$\\
\midrule
  Optimal	& $68.51$ 	& $0.020$ 	\\
 	Sub-optimal & $18.30$   	& $0.025$ 	\\
\bottomrule
\end{tabular}}
\end{center}
\caption{Comparison between the performances of the optimal and the sub-optimal design process.}\label{tab:suboptimal_comparison} 
\end{table}

\section{Optimal DoE applied to SPMe} \label{sec:case_study}
In this section, the optimal DoE described in Sec. \ref{sec:method} is applied  to the SPMe, assuming this latter  to be the real plant \textcolor{black}{and a comparison with  standard identification methods (CC and multistep discharging profiles) is shown. A zero mean Gaussian measurement noise  of $0.3mV$ is considered, with a corresponding variance of  $\sigma_y^2=0.09\cdot10^{-6}$.} Such noise is chosen to be  consistent with the precision of the mostly used experimental instruments. In the following, the output is sampled with  $t_s=5s$. All the proposed approaches have the same initial condition $x_0$   is 
\begin{align}
x_0=\left[3900\,\,14870\,\,0 \,\,\, 0\,\,\, 2000 \,\,\, ... \,\,\, \,\,\, 2000\,\,\, ... \,\,\, \,\,\, 2000\right].
\end{align}
\subsection{Optimal design approach}
The proposed method is based on a sequence of  experiments $\xi_i, \,\, i=1,...\, n$,   each with   duration $t_f^{\xi_i}-t_i^{\xi_i}=1000s$  subdivided, for computational reasons, into $M=4$ sub-optimal steps (Algorithm \ref{alg:suboptimal}). 
\textcolor{black}{The optimal input sequence is piece-wise constant over each $t_s=5s$.}
In the following, the performance of the optimal DoE are evaluated in terms of convergence of the estimated parameter vector to the true value and variances reduction over  $n=10$ experiments. All the experiments present the same duration and initial condition.
After each experiment, \textcolor{black}{ in order to bring the initial condition back to $x_0$ ($x_0^{\xi_i}=x_0,\,\, i=1,\cdots n$),  a CC charging current of $1C$, followed by a resting period of $T_{rest}=400s$, is applied.}
The input sequence of the optimal DoE is obtained as described in Sec. \ref{sec:suboptimal_method}, with
\begin{subequations} 
\begin{align}
-I_{\text{max}} &\leq u(t) \leq I_{\text{max}}\\
V_{\text{min}} &\leq y(t) \leq V_{\text{max}},
\end{align}
\end{subequations}

where  $I_{\text{max}}=1 C$, $V_{\text{min}}=2.5 V$ and $V_{\text{max}}=4.35 V$. 
\subsection{CC discharging approach}
\begin{figure}[htpb]
\begin{center}
\includegraphics[width=0.49\textwidth]{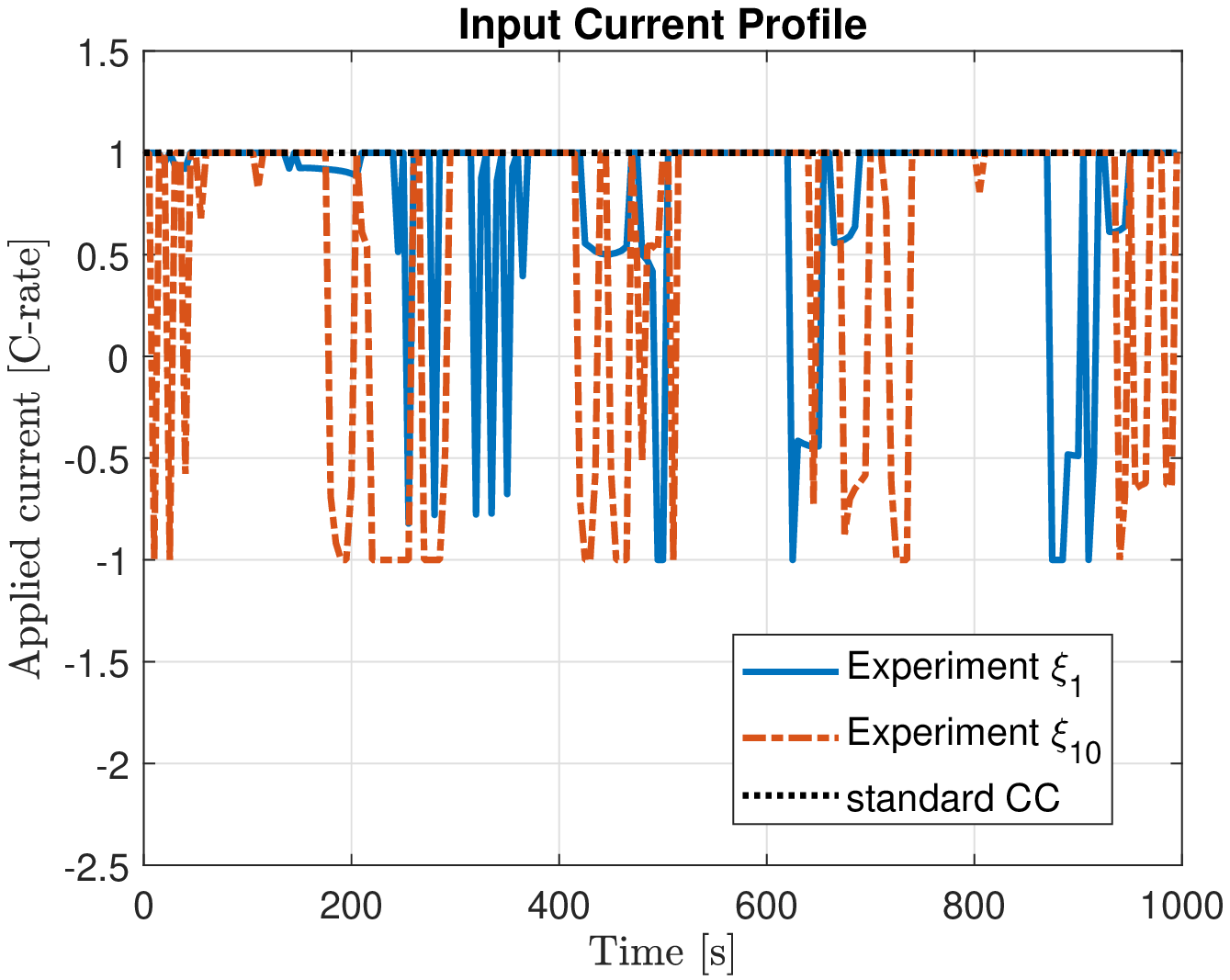}    
\includegraphics[width=0.49\textwidth]{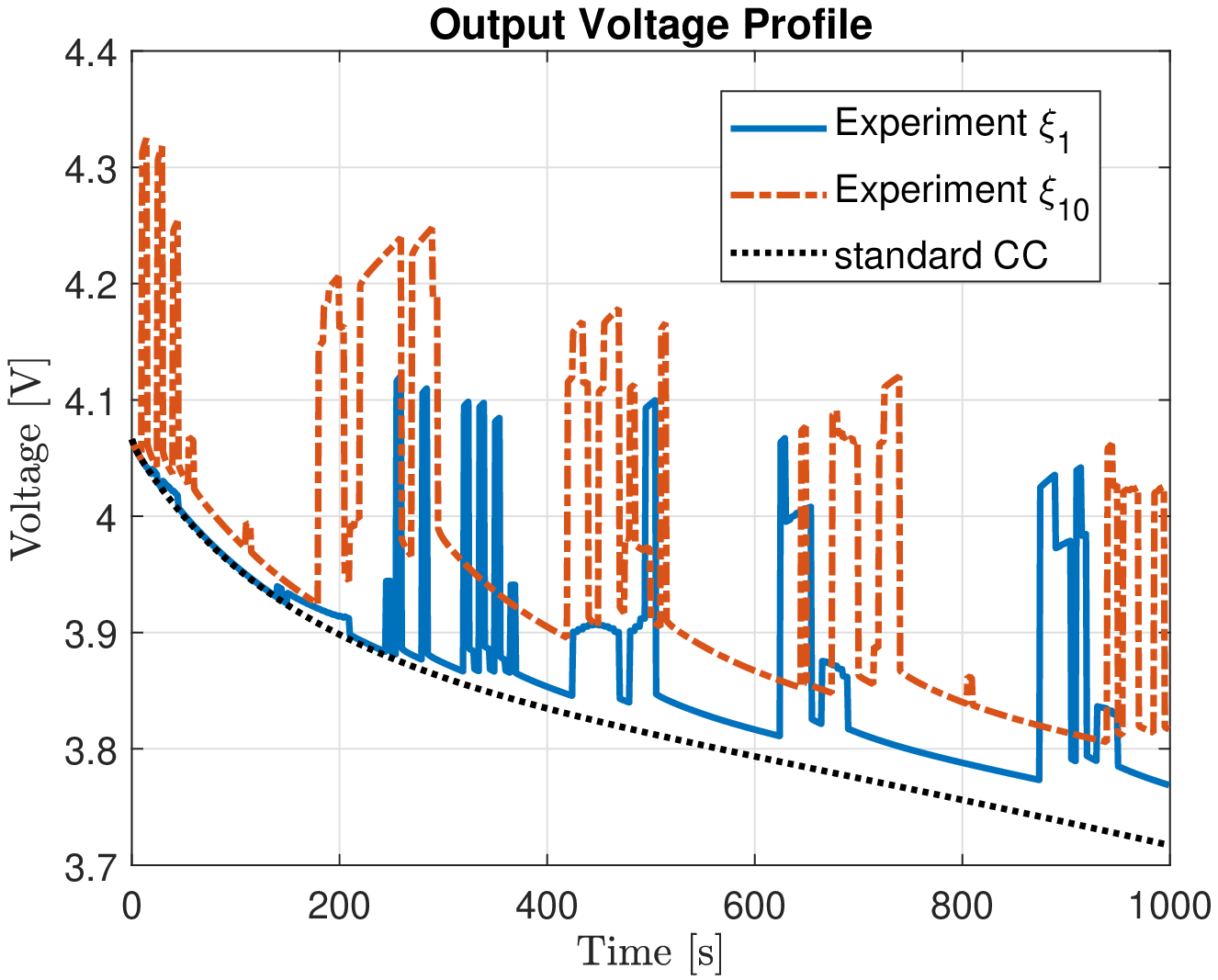}    
\caption{Current input profile and voltage output response applied during the optimal DoE.}  
\label{fig:doe_profile}                                 
\end{center}                                 
\end{figure}

The CC discharging approach is based on a sequence of  experiments $\xi_i, \,\, i=1,...\, n$, with $n=10$, each of them consisting of a simple constant current discharging protocol with $1C$ and time duration $1000s$. \textcolor{black}{Also in this case, at the beginning of each experiment, the initial condition is brought back to $x_0$ ($x_0^{\xi_i}=x_0,\,\, i=1,\cdots n$) by applying a CC charging current of $1C$, followed by a resting period of $T_{rest}=400s$.}
Fig. \ref{fig:doe_profile}  reports the current and voltage profiles  obtained with the optimal DoE and the CC discharging approach. 
\subsection{Multistep discharging approach}
\begin{figure}[htpb]
\begin{center}
\includegraphics[width=0.49\textwidth]{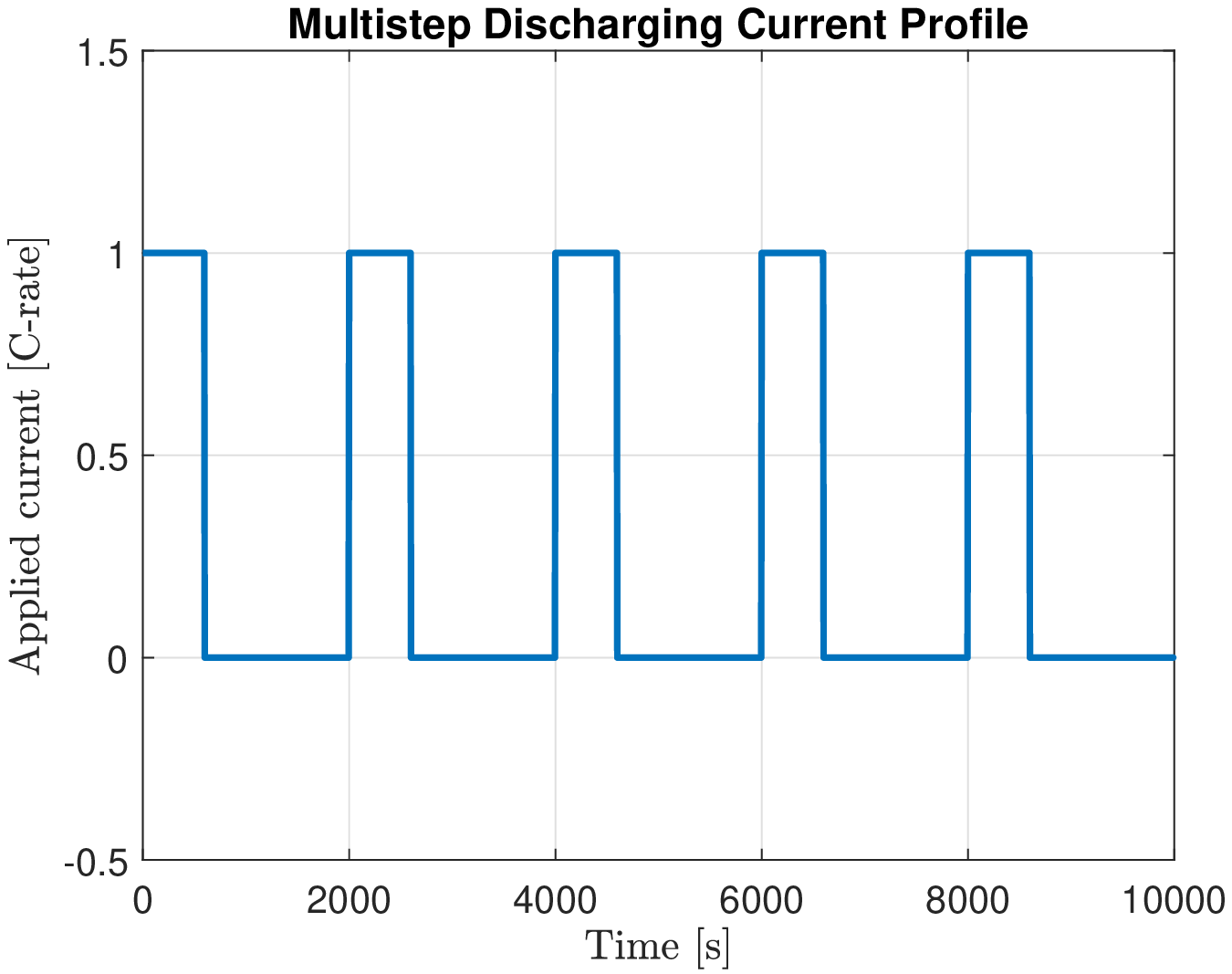}    
\includegraphics[width=0.49\textwidth]{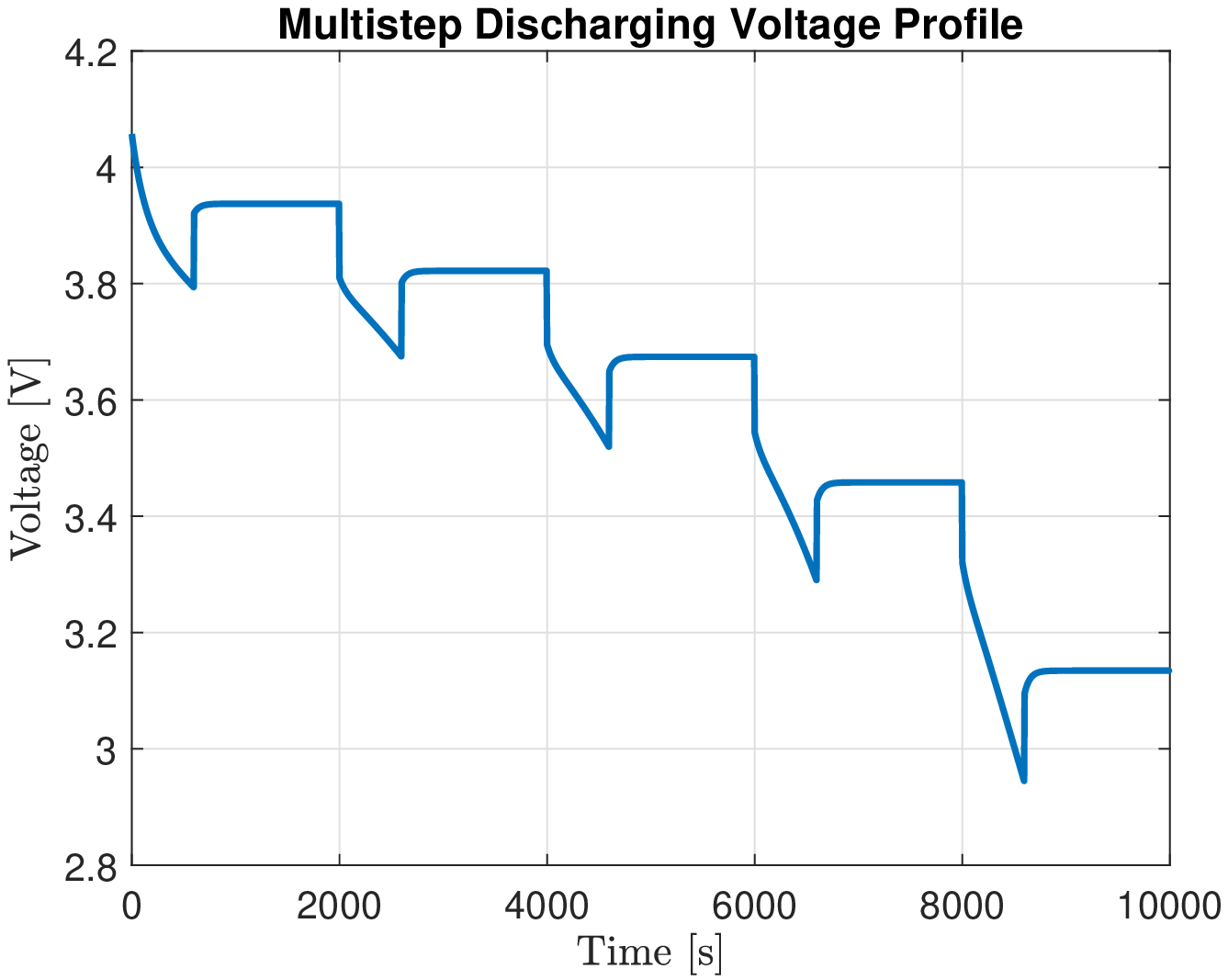}  
\caption{Current and voltage  profiles used for multistep discharge identification.}  
\label{fig:hppc_profile}                                 
\end{center}                                 
\end{figure}

\textcolor{black}{
The  multiple step discharging current profile   explores the whole battery state of charge range, as described in Fig. \ref{fig:hppc_profile}.  In particular, the battery is  discharged from $100\%$ of state of charge  in $5$ steps. Each step applies a $1C$ current for $600s$ followed by  $1400s$ of rest. Differently from the previous approaches the multistep discharging protocol consists of a single experiment of time duration $10000s$. As it can be noticed the overall duration of all the approaches is the same ($10000s$).}

\subsection{Results comparison}
\begin{figure}[htpb]
\begin{center}
\includegraphics[width=0.5\textwidth]{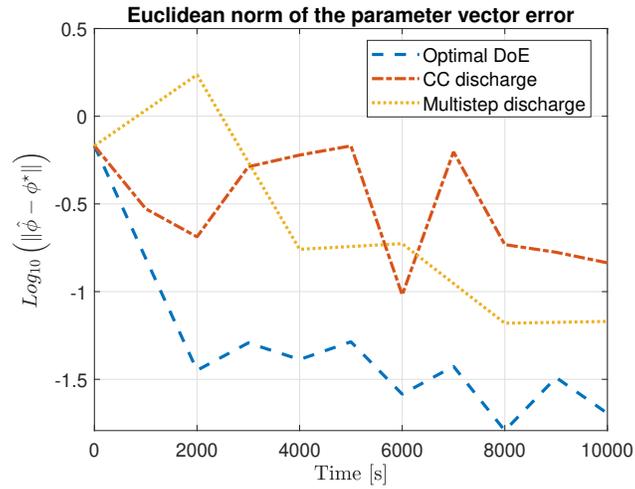}    
\caption{Comparison between the optimal experimental approach and the standard ones \textcolor{black}{in terms of Euclidean distance of the parameter vector to the true value.}}  
\label{fig:accuracy_compare}                                 
\end{center}                                 
\end{figure}

\begin{figure}[htpb]
\begin{center}
\includegraphics[width=0.5\textwidth]{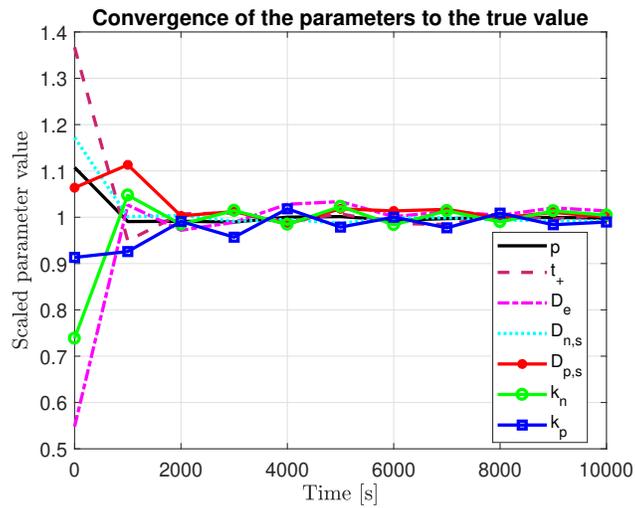}    
\caption{Convergence of the parameters to the true values.}  
\label{fig:convergence_parameter}                                 
\end{center}                                 
\end{figure}

In Figure \ref{fig:accuracy_compare}, the  experimental procedures are compared in terms of convergence  of the scaled estimated parameter vector $\hat{\phi}$ to the  true values $\phi^\star$, i.e. \textcolor{black}{in terms of the euclidean norm 
\begin{align}
\Vert \hat{\phi}-\phi^* \Vert=\sqrt{(\hat{\phi}-\phi^*)^T(\hat{\phi}-\phi^*)}.
\end{align}} 
\textcolor{black}{As it can be noticed, the parameter vector identified using the optimal DoE presents a fast convergence to the real value and a very low error even after the first experiment ($1000s$). On the other side, standard approaches require a longer time to reduce the gap between the identified parameter vector and the true one. }
\textcolor{black}{
Figure \ref{fig:convergence_parameter} highlights how the different SPMe scaled parameters convergence to  their true values using the optimal DoE approach. }

\textcolor{black}{
Table \ref{tab:par_estimated}  shows the parameter vector estimated after the last experiment for all the  approaches. In particular, the optimal DoE results the most accurate, although all the approaches show low estimation error  at the end of the experimental procedure. Note that, thanks to the behavior shown in Figure \ref{fig:accuracy_compare}, the optimal DoE already guarantees low estimation error after $1000s$.}
\begin{table} [htpb] 
\begin{center}
{\renewcommand\arraystretch{1.4} 
\begin{tabular}{*6c} 
\toprule
Parameter  						& \textcolor{black}{$\phi_0$}			& $\phi^*$			 	& $\hat{\phi}_{DoE}$  & $\hat{\phi}_{CC}$ & $\hat{\phi}_{m-step}$ 	\\
\midrule
$p$ 							&\textcolor{black}{$1.1075$}				&$1	$					& $0.9979$   		& 	$1.0105$ 		& \textcolor{black}{$0.9911$}	\\
$t_+$ 							&\textcolor{black}{$1.3668$}			&$1	$																				& $1.0052$ 			&	$0.9944$ & \textcolor{black}{$0.9579$	} 	\\
$D_e$ 							&\textcolor{black}{$0.5482$}		 	&$1	$																					& $1.0136$   & $1.0975$ & \textcolor{black}{$0.9990$}	\\
$D_{s,p}$ 						&\textcolor{black}{$1.1724$}		&$1	$																			& $0.9920$   & $0.9707$  & \textcolor{black}{$1.0029$	}	\\
$D_{s,n}$ 						&\textcolor{black}{$1.0638$}			&$1	$																				& $1.0004$  & $1.0192$ & \textcolor{black}{$1.0498$	} \\
$k_p$ 							&\textcolor{black}{$0.7385$}			&$1	$																& $1.0044$ & $1.0866	$	& \textcolor{black}{$0.9903$} \\
$k_n$ 							&\textcolor{black}{$0.9133$}			&$1	$									& $0.9896$ & $0.9460$ 								& \textcolor{black}{$0.9885$	}			 \\
\bottomrule
\end{tabular}}
\end{center}
\caption{\textcolor{black}{Identified scaled parameters at the end of the experiment procedure using both optimal and standard approaches compared with the initial guess and the true value.}}\label{tab:par_estimated}
\end{table}
Considering the indexes introduced in Sec. \ref{sec:ill_conditioning_sources} one can  evaluate the presence of ill-conditioning sources and unidentifiable parameters. In particular, the value of such indexes after the last experiment are reported in Table \ref{tab:indexes}. \textcolor{black}{As it can be noticed, the  collinearity index and the  condition number are higher using the CC discharging approach than the optimal DoE and the mulstistep. This implies that, using the CC discharging approach, some parameters may remain  unidentifiable after the last experiment. Note that, thanks to the behavior shown in Figure \ref{fig:accuracy_compare}, with the optimal DoE these indexes were already significantly low after $1000s$. }
\begin{table} [htpb] 
\begin{center}
{\renewcommand\arraystretch{1.3} 
\begin{tabular}{*4c} 
\toprule
Index  								& Optimal DoE				& Standard  & Multistep discharge\\
\midrule
$\kappa$ 		&$577$						& $31963$	& \textcolor{black}{$260$}\\		
$\gamma$ 		&$152$					& $7548$	& \textcolor{black}{$110$}\\	
\bottomrule
\end{tabular}}
\end{center}
\caption{Condition number and collinearity index at the end of the experimental realization.}\label{tab:indexes}
\end{table}
Finally, in  Table \ref{tab:variance_comparison} shows  the variance reduction during the experiments realization. \textcolor{black}{ In particular it emerges that the variance of all the parameters  in the optimal DoE   is very low already after the first experiment ($1000s$), while  the multistep discharging approach  requires $10000s$ to achieve high performance. In  the CC discharging method the parameter variance  still remains high at the last experiment.} These results are promising since they underline that an optimal design of experiment can significantly improve parameters estimation  accuracy (both in terms of converging parameter values and variance). Furthermore,  the number of experiments required for the convergence of the parameters to the true values, \textcolor{black}{ i.e. the time-duration of the experimental realization required for an accurate identification,} is much shorter using the optimal DoE  than the standard current profiles.
\begin{table} [htpb] 
\begin{center}
{\renewcommand\arraystretch{1.3} 
\begin{adjustbox}{width={\textwidth},totalheight={\textheight},keepaspectratio}
\begin{tabular}{c|*3c|*3c} 
\toprule
\multirow{2}{*}{Parameter}  			& \multicolumn{3}{c|}{Variance after $\xi_{1}$}    		 & \multicolumn{3}{|c}{Variance after $\xi_{10}$}   \\
\cline{2-7}
 & Optimal DoE & CC discharge &\textcolor{black}{ Multistep discharge }  & Optimal DoE & CC discharge &\textcolor{black}{ Multistep discharge  }\\
\midrule
$p$ 				& $5 \cdot 10^{-5}$ & $1.21$ 		&\textcolor{black}{$0.0636$} & $10^{-5}$ 							& $0.24$ & \textcolor{black}{$10^{-4}$}\\
$t_+$ 				& $8 \cdot 10^{-4}$ & $0.92$ 			& \textcolor{black}{$0.0225$} & $1.6 \cdot 10^{-4}$			 				& $0.19$ & \textcolor{black}{$4 \cdot 10^{-4}$}	\\
$D_e$ 				& $6 \cdot 10^{-3} $ & $6.97$ 			& \textcolor{black}{$0.5488$ }& $1.2 \cdot 10^{-3}$ 							& $1.39$  & \textcolor{black}{$6 \cdot 10^{-4}$}\\
$D_{s,p}$			& $6 \cdot 10^{-4}$ & $0.59$ 			& \textcolor{black}{$0.038$} & $10^{-4}$							&$0.12$  &\textcolor{black}{$10^{-5}$} \\
$D_{s,n}$			& $9 \cdot10^{-4}$ & $0.01$ 			& \textcolor{black}{$0.0166$	}& $1.86 \cdot 10^{-4}$							& $2.9 \cdot 10^{-3}$  & \textcolor{black}{$5 \cdot 10^{-4}$}\\
$k_p$				& $1.5 \cdot 10^{-3}$ 	& $14.88$ 		& \textcolor{black}{$0.8118$ }& $3 \cdot 10^{-4}$ 							&$2.98$ &\textcolor{black}{$10^{-4}$}  \\
$k_n$ 				& $2.5 \cdot 10^{-3}$ 	& $1.10$ 			&\textcolor{black}{ $0.1004$ }& $4.9 \cdot 10^{-4}$ 							&$0.22$ &\textcolor{black}{$10^{-4}$}  \\
\bottomrule
\end{tabular}
\end{adjustbox}}
\end{center}
\caption{Comparison between scaled parameter variances during the identification process, both in case of optimal design of experiment  and standard approach.}\label{tab:variance_comparison}
\end{table}

The average time  needed to solve the optimization problem described in Sec. \ref{sec:suboptimal_method} and compute the current profile for a single experiment is  $520.81\,s$, while the average time  needed to the solve identification process in \eqref{eq:est}  is $13.59\,s$. 
The simulations are performed on a Windows 10 machine with 16Gbytes of RAM and Intel core I7-6700HK quad core processor \@3.5 GHz. Note that,  since the design of experiment and the parameter estimation process  can be conducted offline, the time required by the proposed method appears suitable for real applications.

\section{Optimal DoE applied to P2D model}
\label{sec:p2d_esperiment}
In the section above, the optimal DoE has been conducted  on the SPMe, assuming that the model used as the real plant and the one used for the parameters estimation were the same. For this reason, the measurement data has been  collected by simulating the SPMe output affected by a zero mean gaussian error. However, while the SPMe is particularly suitable for control purposes, its use for accurately simulating a real lithium-ion cell   may be inadequate. For this reason, in the following, \textcolor{black}{LIONSIMBA, a battery simulator which  implements} the  P2D model, is assumed to be the real plant, with  the output voltage affected   by a zero mean gaussian error of $0.3mV$, with variance $\sigma_y= 0.09^{-6}$.  
\textcolor{black}{It is important to notice that the use of two models, a model for simulation (very detailed) and a model for control (simpler), is a well known procedure. In particular, this allows to assess, in a preliminary way, the practical effectiveness of a novel control approach thus reducing time and costs during the experimental phase}. 

\begin{figure}[htpb]
\begin{center}
\includegraphics[width=0.5\textwidth]{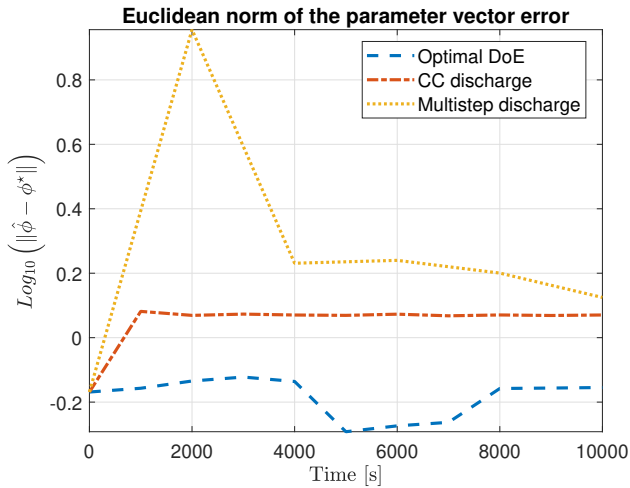}    
\caption{Comparison between the optimal DoE  and the standard approach, experimentally applied to \textit{LIONSIMBA}.}  
\label{fig:accuracy_p2d}                                 
\end{center}                                 
\end{figure}

In the following, we assign value $\phi^*$  (see Table \ref{tab:parameters_estimated}) to the parameters of the P2D. Note that, in general, the parameter vector resulting from the optimization may be different from $\phi^*$. In fact, although physically meaningful,  the use of a simplified model  with  parameter values equal to the ones of the real plant (P2D) may not be the best choice in terms of output fitting. In particular, the SPMe implementation with the parameter vector $\phi^*$ presents  an RMS error in the voltage prediction of the P2D during normal cycling around few $mV$. This lack of accuracy comes in the form of a bias, and, although this  could seem negligible, such error may cause problems in the context of state estimation.   In the following, the parameters of the SPMe with the best fitting in terms of voltage are  considered unknown, as in real experiments, and the parameter vector $\phi^*$ in Table \ref{tab:par_estimated} is used only to simulate for comparative purposes the SPMe with the P2D parameters.  Furthermore, the optimal DoE is compared with standard experimental approaches, such as  CC and multistep discharging protocols. The experimental setting adopted is the same of  Sec. \ref{sec:case_study}. The experiments are performed on \textit{LIONSIMBA} \cite{torchio2016_LIONSIMBA}).
Figure \ref{fig:accuracy_p2d} shows  the  Euclidean distance between the estimated parameter vector $\hat{\phi}$ and $\phi^*$.
As discussed above, such distance is not expected to go to zero, since it is not guaranteed  that  the best fit will come with  $\phi^*$.  However, a low value of this Euclidean norm ensures   a physical meaning of the estimated parameters. \textcolor{black}{Note that, in real experiments the true parameters are unknown but we can evaluate the model performance in terms of output fitting of validation data and in terms of closedness  to the data-sheet parameters. }
 As it can be noticed from Figure \ref{fig:accuracy_p2d}, the parameter vector identified with the optimal DoE is closer  to the one of the P2D model than in the standard approaches.  
Table \ref{tab:par_estimated_p2d}  shows the parameters estimated after the last experiment for all the  approaches. In this case,   many parameters such as the  Bruggeman coefficient (using the CC discharging approach) and the electrolyte diffusion coefficient (using the multistep discharging approach)  seem losing their physical meaning (in bold, in Table \ref{tab:par_estimated_p2d}), while the optimal DoE allows to identify parameters close to the nominal value.  
\begin{table} [htpb] 
\begin{center}
{\renewcommand\arraystretch{1.4} 
\begin{tabular}{*5c} 
\toprule
Parameter  								& $\phi^*$				& $\hat{\phi}_{DoE}$ 	& $\hat{\phi}_{cc}$ & $\hat{\phi}_{m-step}$ \\
\midrule
$p$ 									&$1	$									 											& $0.6150$   	& 	$0.1000$	& 	\textcolor{black}{$ 0.8776$}			\\
$t_+$ 									&$1	$																				& $0.9472$ 		&	$0.8537$	& \textcolor{black}{$0.8905$}	 	\\
$D_e$ 								 	&$1	$																					& $0.4478$   & $0.4303$ 	&\textcolor{black}{$2.2264$}\\
$D_{s,p}$ 								&$1	$									 											& $1.1164$   	& $0.8236$ &\textcolor{black}{$0.8508$}\\
$D_{s,n}$ 								&$1	$																				& $0.9065$  & $0.7288$ &\textcolor{black}{$0.7400$	} \\
$k_p$ 									&$1	$													& $1.0011$ & $0.7331$				& \textcolor{black}{$1.3804$ }\\
$k_n$ 									&$1	$																				& $0.8897$  & $0.7742$ &\textcolor{black}{$ 1.1032$}	\\
\bottomrule
\end{tabular}}
\end{center}
\caption{Identified scaled parameter of the SPMe after 10 experiments conducted on the P2D model.}\label{tab:par_estimated_p2d}
\end{table}
\subsection{Validation of the SPMe with the identified parameters}
\begin{figure}[htpb]
\begin{center}
\includegraphics[width=0.5\textwidth]{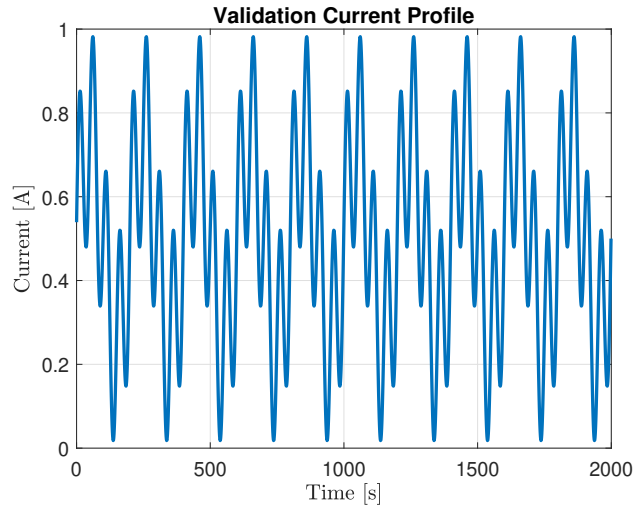}    
\caption{Input current profile used for validation.}  
\label{fig:validation_profile}                                 
\end{center}                                 
\end{figure}

\begin{figure}[htpb]
\begin{center}
\includegraphics[width=0.5\textwidth]{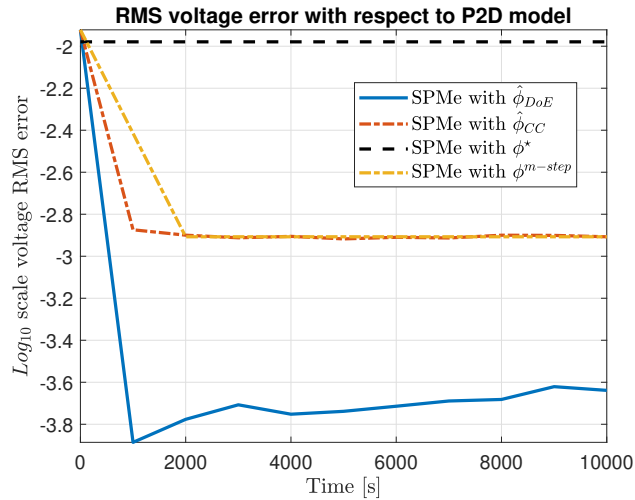}    
\caption{Comparison between the optimal experimental approach and the standard ones in terms of  root mean square error.}  
\label{fig:ERR_rms_p2d}                                 
\end{center}                                 
\end{figure}

In this section, the SPMe is validated, in terms of P2D voltage prediction, using the  parameters  identified in the section above with the optimal DoE and the standard approaches. Furthermore, the comparison with the SPMe  using  $\phi^*$ is shown. \textcolor{black}{In particular, the  current input profile $u_{val}(t)$ used for the validation (Figure \ref{fig:validation_profile}) consists of a biased multi-sinusoidal  current with frequency $f_1=20\,\text{mHz}$ and $f_2=5\,\text{mHz}$,  $I_{bias}=0.5\,\text{C}$   as mean value and $I_{sin}=0.25\,\text{C}$  as peak value of the sinusoidal components
\begin{align}
u_{val}(t)=I_{sin}\left(\sin(2\pi f_1 t)+ \sin(2\pi f_2 t)\right)+ I_{bias}.
\end{align}
}
\textcolor{black}{
The results of the validation process are shown in Figure \ref{fig:ERR_rms_p2d}. This  latter reports the evolution of the voltage RMS  error between the SPMe  and the P2D, evaluated after each experiment on the validation profile.  As it can be noticed, the RMS error  using the optimal DoE is very low after the first experiment, while the standard approaches present a slow convergence to an higher error. Note that, all the estimation methods improve the accuracy in the P2D voltage prediction given by the SPMe with $\phi^*$.  On the other side,  the optimal DoE is the one which provides  physically meaningful parameters in the shortest time. }  

The validation performed so far shows  promising results in the application of the optimal DoE to the P2D model, used in this context as the real plant. In particular, it is demonstrated that the estimation of the SPMe parameters according to the optimal DoE can significantly increase the model performance in the P2D voltage prediction.

\section{Conclusion} \label{sec:conclusions}

The use of accurate models in \textcolor{black}{advanced BMSs} is necessary in order to achieve high performance in battery operations. For this reason, a suitable identification process is required. Note that, the input signal  has to be sufficiently exciting during the experiment realization, in order to  reduce parameters uncertainty.  In this work, the optimal DoE is applied to the SPMe, so to maximize the parameters accuracy and the results are compared with standard identification profiles, such as CC and \textcolor{black}{multistep discharging approaches. A sub-optimal approach is proposed in order to reduce the computational burden, which  may be a limiting factor in the design of   experiments of long time duration, which plays a key role in  parameters identification accuracy}.  \textcolor{black}{The results show that the proposed methodology outperforms standard approaches  in terms of  time required for the convergence  of the parameters  when the SPMe is assumed to be the real plant.} Subsequently, the P2D model is considered as the real plant and the SPMe used as model for the control. Also in this case,   the optimal experimental design for identifying the parameters of the SPMe provides the best results during  validation. 
\textcolor{black}{Future works may include experimental validation of the proposed strategy}. 


\section{Supporting information}
\label{sec:supporting}
In the Supporting Information file we include the tables with the parameters of the P2D model used in the simulations, in order to increase their reproducibility.

\bibliography{refs_battery}

\end{document}